\documentclass[11pt, a4paper]{article}

\usepackage[top=2cm, bottom=2.5cm, left=2cm, right=2cm]{geometry}
\pdfoutput=1 
\usepackage{jcappub}
\usepackage{graphicx}
\usepackage{amsmath}
\usepackage{amsfonts}
\usepackage{amssymb}
\usepackage{fancyhdr}
\usepackage{epic}
\usepackage{setspace}
\usepackage{multirow}
\usepackage{lastpage}
\usepackage{array}
\usepackage{hhline}
\usepackage{booktabs}
\usepackage{wrapfig}
\usepackage{graphicx,bm,color}
\usepackage{cancel}
\usepackage{slashed}
\usepackage{hyperref}
\usepackage{pstricks}
\usepackage{color}
\usepackage{tabularx,ragged2e}
\usepackage{eufrak}
\usepackage{subcaption}

\DeclareMathOperator{\sign}{sign}

\title{\LARGE  Baryogenesis via gauge field production from a \\ relaxing Higgs}

\date{\today}

\author[a]{Yann Cado,}
\author[a]{Benedict von Harling,} 
\author[a,b]{Eduard Mass\'o,} 
\author[a]{\\ Mariano Quir\'os} 

\affiliation[a]{
Institut de F\'isica d'Altes Energies (IFAE), The Barcelona Institute of Science and Technology, Campus UAB, 08193 Bellaterra (Barcelona), Spain}
\affiliation[b]{Grup de F{\'\i}sica Te{\`o}rica, Departament de F{\'\i}sica, Universitat Aut{\`o}noma de Barcelona, 08193 Bellaterra, Spain}
\emailAdd{ycado@ifae.es}
\emailAdd{bvonharling@ifae.es}
\emailAdd{eduard.masso@ifae.es}
\emailAdd{quiros@ifae.es}

\abstract{
We show that the baryon asymmetry of the universe can be explained in models
where the Higgs couples to the Chern-Simons term of the hypercharge group
and is away from the late-time minimum of its potential during inflation. The Higgs then relaxes toward this minimum once inflation ends which leads to the production of (hyper)magnetic helicity. We discuss the conditions under which this helicity can be approximately conserved during its joint evolution with the thermal plasma. At the electroweak phase transition the helicity is then converted into a baryon asymmetry by virtue of the chiral anomaly in the standard model. We propose a simple model which realizes this mechanism and show that the observed baryon asymmetry of the universe can be reproduced. 
}


\def \be {\begin{equation}}
\def \ee {\end{equation}}
\def \bi {\begin{enumerate}}
\def \ei {\end{enumerate}}
\def \ba {\begin{aligned}}
\def \ea {\end{aligned}}
\def \bse {\begin{subequations} \begin{eqnarray}}
\def \ese {\end{eqnarray} \end{subequations}}

\def \A {\bm{A}}

\def \B{\bm{B}}
\def \AB {\A \cdot \B}

\renewcommand{\exp}[1]{\text{exp} \left(#1 \right)}
\newcommand{\moy}[1]{\langle #1 \rangle}

\newcommand{\pder}[2]{\dfrac{\partial #1}{\partial #2}}

\DeclareMathOperator{\diag}{diag}

\allowdisplaybreaks

\begin{document}
\maketitle

\section{Introduction}
\label{sec:introduction}

Electroweak (EW) baryogenesis is an appealing mechanism which can in principle generate the baryon asymmetry of the universe during the EW phase transition. However, it requires this phase transition to be first-order, while in the standard model (SM) it is a smooth crossover. New physics beyond the SM is then necessary to make it first-order. In this paper, we will consider a scenario in which the baryon asymmetry is also generated during the EW phase transition but the latter can be a crossover like in the SM. It has another similarity with EW baryogenesis in the sense that the dynamics of the Higgs is important for the mechanism.  Compared to EW baryogenesis though, part of this dynamics takes place at much earlier times, after the end of inflation. 

We assume that the Higgs $\Phi$ couples to the Chern-Simons term of the hypercharge gauge group, $ |\Phi|^2 Y_{\mu \nu} \tilde{Y}^{\mu \nu}$. If the Higgs is elongated away from the late-time minimum of its potential during inflation, it will start to relax toward this minimum after inflation ends. During this stage, the EW symmetry is broken. The evolution of the Higgs results in the copious production of photons via the above coupling. As we will see, these photons are predominantly produced with one helicity, leading to a helical background of photons. Once the universe reheats and the EW symmetry is restored, they are transformed into helical hypermagnetic fields.  The latter subsequently start to interact and evolve jointly with the thermal plasma. Under certain conditions, which we will discuss in some detail, the helical hypermagnetic fields can survive until the EW phase transition. The EW symmetry is then once again broken and the hypermagnetic fields are transformed into ordinary magnetic fields. The hypercharge gauge boson contributes to the anomaly of baryon plus lepton number, $B+L$, and changes in the hypermagnetic helicity therefore result in changes of the $B+L$-charge \cite{Giovannini:1997eg}. Since the photon does not contribute to the anomaly, the transformation of the hypermagnetic fields into ordinary magnetic fields leads to a compensating $B+L$-asymmetry.  While it is being generated, EW sphalerons start to erase this asymmetry but they freeze out toward the end of the phase transition when the source for the asymmetry is still active. By carefully analyzing the dynamics of the EW phase transition in the SM, it was shown in \cite{Kamada:2016cnb} that a sizeable $B+L$-asymmetry can survive which can reproduce the observed baryon asymmetry of the universe.

This baryogenesis mechanism has previously been considered in the context of axion inflation, where the hypermagnetic helicity is produced during inflation via the coupling of the axion-like inflaton to the Chern-Simons term of the hypercharge gauge group \cite{Anber:2015yca,Cado:2016kdp,Jimenez:2017cdr,Domcke:2019mnd}. Other related scenarios involving the decay of hypermagnetic helicity to generate the baryon asymmetry have been studied in \cite{Brustein:1998du,Brustein:1999rk,Giovannini:1999wv,Giovannini:1999by,Bamba:2006km,Kamada:2016eeb,Kamada:2018tcs,Barrie:2020kpt}.  
The relaxation of the Higgs toward the minimum of its potential after inflation, on the other hand, has been used in \cite{Kusenko:2014lra,Pearce:2015nga,Yang:2015ida} to implement spontaneous baryogenesis via a coupling of the Higgs to the $B+L$-current. This is related to our scenario by a chiral rotation of the SM fermions (but also differs in important aspects).

To induce a large vacuum expectation value (VEV) for the Higgs during inflation, we consider a coupling to the Ricci scalar. The sign of this coupling is chosen such that it leads to a tachyonic mass term for the Higgs.  This mass term and thus the minimum of the Higgs potential are approximately constant during inflation but they decrease rapidly once inflation ends. The Higgs then starts moving from its  large initial VEV toward the origin of its potential. This leads to the production of a helical background of photons as mentioned before. Let us emphasize that the coupling to the Ricci scalar is only one option to obtain a large initial VEV for the Higgs. It has the advantage that it allows us to consider only the time evolution of the Higgs. We also comment on another option, a coupling of the Higgs to the inflaton which also results in a tachyonic mass term, but leave a more detailed study to future work.

As we will discuss, the magnetic fields which are produced after inflation need to be sufficiently strong to guarantee their survival until the EW phase transition via a process known as the inverse cascade. We find that this in turn means that the initial Higgs VEV needs to be quite large, $\gg 10^{13}\,$GeV. At such large VEVs, the Higgs quartic coupling is expected to run to negative values in the SM. This is dangerous since the Higgs could be driven into the resulting unphysical deeper minimum during inflation. In order to avoid this, we will couple the Higgs to a scalar singlet whose quantum corrections ensure that the Higgs quartic coupling stays always positive. Again, this is only one option and of course different new physics could be introduced to achieve the same effect. Armed with a model for the Higgs potential, we numerically solve for the time evolution of the Higgs and from this calculate the produced helicity and magnetic field strength. For definiteness, we study three benchmark points which span a representative set of possibilities. Our main result is that the helicity can survive until the EW phase transition and that its conversion can reproduce the observed baryon asymmetry of the universe. 

This paper is organized as follows. We derive the relevant equations of motion (EOMs) for the Higgs and the photon in sec.~\ref{sec:model} and comment on the backreaction of photon production on the Higgs, finite-temperature effects and the simultaneous generation of asymmetries in SM fermions. In sec.~\ref{sec:HelicityEvolution}, we then discuss the evolution of the helicity and the asymmetries between reheating and the EW phase transition and derive conditions on the survival of the helicity. Furthermore, we estimate the baryon asymmetry that is generated during the EW phase transition from the conversion of the helicity. We specify the Higgs potential in sec.~\ref{sec:result} and present our numerical results for the three benchmark points. Finally, we conclude in sec.~\ref{sec:conclusions}. Three appendices complement the paper. In appendix~\ref{appendix:UVcompletion}, we give an example of a ultraviolet (UV) completion for the required coupling of the Higgs to the hypercharge gauge boson. The renormalization group (RG) equations for the model with the added singlet scalar are presented in appendix~\ref{appendix:RGEs} and the initial conditions of the Higgs after inflation are derived in appendix~\ref{appendix:HiggsPotentialInflation}.

\section{Helicity from a relaxing Higgs}
\label{sec:model}

\subsection{Higgs equation of motion}

We consider the action
\be
\label{eq:action}
S \, = \, \int d^4 x \left[ \sqrt{-g} \left(- g^{\mu \nu } D_{\mu} \Phi D_\nu \Phi^\dagger \, - \, V(\Phi) \, -\, \frac{1}{4} g^{\mu \nu } g^{\rho \sigma } Y_{\mu \rho} Y_{\nu \sigma}  \right)  \, + \, \frac{1}{2}  \frac{|\Phi|^2}{M^2} Y_{\mu \nu} \tilde{Y}^{\mu \nu} \right]  ,
\ee
where $\Phi$ is the Higgs doublet, $V$ its potential, $Y_{\mu \nu}$ the field strength of the hypercharge gauge field $A_{Y \mu}$ and  $\tilde{Y}^{\mu \nu}= \epsilon^{\mu \nu \rho \sigma} Y_{\rho \sigma} / 2$ with $\epsilon^{0123}=1$. We will use the conformally-flat Robertson-Walker metric $g_{\mu \nu} = a^2 \,\eta_{\mu \nu}$, where $a$ is the scale factor and $\eta^{\mu \nu}= \diag(-1,1,1,1)$. The details of the UV completion that gives rise to the higher-dimensional coupling between the Higgs and the hypercharge gauge boson will not be important. One example of such a UV completion is presented in appendix \ref{appendix:UVcompletion}. The mass scale $M$ will be specified later.

We will be interested in the relaxation of the Higgs from some initial value after inflation to the minimum of its potential. The VEV of the Higgs is then large (except for the brief moments when it crosses zero in case it oscillates) and EW symmetry is broken. The higher-dimensional coupling in eq.~\eqref{eq:action} is accordingly replaced by couplings to the photon and the $Z$ boson. Since the $Z$ boson is massive, we will focus on the coupling to the photon. Furthermore, all but one degree of freedom of the Higgs doublet are eaten and we get
\be
\label{eq:LEaction}
S \, = \, \int d^4 x \left[  - \, \frac{1}{2} \, a^2 \partial_{\mu} h \, \partial^\mu h  \, - \, a^4 \, V(h)  \, - \, \frac{1}{4} F_{\mu \nu} F^{\mu \nu} \,+ \, \frac{\cos^2 \theta_W}{4}  \frac{h^2}{M^2} F_{\mu \nu} \tilde{F}^{\mu \nu} \, + \, \dots\right] ,
\ee
where $h$ is the real Higgs field, $F_{\mu \nu}$ the field strength of the photon $A_\mu$ and $\theta_W$ the EW angle.  
From here onwards, all spacetime indices are contracted with $\eta^{\mu \nu}$.

We will consider the evolution of the Higgs and the photon starting at the end of inflation. We assume that the inflaton initially oscillates in its potential, leading to a matter-dominated phase before reheating. Furthermore, we assume that perturbative and non-perturbative decays of the inflaton during this period are negligible and set the temperature to zero. The validity of the latter assumption will be discussed in sec.~\ref{sec:backreaction}. The EOMs for the Higgs and the photon read
\begin{gather}
\label{HiggsEOM}
\Box \, h \, - \, \frac{4}{\tau} \frac{\partial}{\partial \tau} h \, - \, \frac{\tau^4}{\tau_{\rm md}^4} \frac{d}{d h} V(h) \, +\,     \frac{\cos^2 \theta_W }{2}\frac{\tau_{\rm md}^4}{\tau^4} \frac{h}{M^2} \,  F_{\mu \nu} \tilde{F}^{\mu \nu} \, = \, 0\\
\label{PhotonEOM}
\partial_\mu F^{\mu \nu} \, - \, \cos^2 \theta_W \,  \epsilon^{\mu \nu \rho \sigma}  \frac{\partial_\mu h^2}{M^2} \partial_\rho A_\sigma \, = \, 0 \, ,
\end{gather}
where $\tau$ is the conformal time.
Here and below we fix $a=1$ at the onset of matter domination which corresponds to the conformal time $\tau_{\rm md} = 2/H_{\rm inf}$, where $H_{\rm inf}$ is the Hubble rate at the end of inflation.

If the Higgs is away from its late-time minimum after inflation, it subsequently relaxes toward this minimum. As we will momentarily see, the coupling to the Higgs in eq.~\eqref{PhotonEOM} then leads to the production of photons. The latter can in turn backreact on the Higgs via the coupling in eq.~\eqref{HiggsEOM}. As discussed in \cite{Adshead:2015pva,Adshead:2016iae}, if this backreaction is important, it leads to the excitation of higher-momentum modes of the Higgs and it is no longer sufficient to consider the evolution of the zero mode. 
In our numerical simulations, we will always ensure that the backreaction is sufficiently small to be neglected which will yield a condition on the ratio $h/M$ at the end of inflation. We can then restrict ourselves to the zero mode of the Higgs, {\it i.e.} $h(\tau,\bm{x}) = h(\tau)$, and neglect the coupling to the gauge field in its EOM which simplifies to
\be
\label{HiggsEOMs}
\frac{\partial^2}{\partial \tau^2}\, h \, + \, \frac{4}{\tau} \frac{\partial}{\partial \tau} h \, + \, \frac{\tau^4}{\tau_{\rm md}^4} \frac{d}{d h} V(h) \, =\, 0 \, .
\ee
For a given potential and initial condition, we can now solve for the time evolution of the Higgs. 

\subsection{Production of gauge fields and fermionic asymmetries}
\label{sec:GaugeFieldProduction}

In order to see how the relaxing Higgs leads to the production of photons, we next quantize the photon. We define $A_{\mu} = (A_{0},\bm{A})$ and work in radiation gauge\footnote{To be more precise, we choose Coulomb gauge $\bm{\nabla} \cdot \bm{A} =0$. The $\nu=0$-component of eq.~\eqref{PhotonEOM} together with $\bm{\nabla} h(\tau)=0$ then allows us to set $A_0=0$.} $A_0=0$ and $\bm{\nabla} \cdot \bm{A} =0$. Going to momentum space, we then get
\be
\bm{A}(\tau , \bm{x}) \, = \, \sum_{\lambda = \pm} \int \frac{d^3 k}{(2\pi)^3} \, \left [\bm{\epsilon}_\lambda(\bm{k}) \, a_\lambda(\bm{k}) \, A_\lambda(\tau, \bm{k}) \, e^{i \bm{k} \cdot \bm{x}} + \, \text{h.c.} \right]  , 
\ee
where $\lambda = \pm$ is the helicity of the photon and the $a_\lambda(\bm{k})$ are annihilation operators that fulfill the canonical commutation relations. The polarization vectors satisfy
\be
\bm{k} \cdot \bm{\epsilon}_\lambda(\bm{k}) = 0\, ,  \quad \bm{k} \times \bm{\epsilon}_\lambda(\bm{k})  = - i \lambda k \, \bm{\epsilon}_\lambda(\bm{k})\, , \quad \bm{\epsilon}^*_{\lambda'}(\bm{k}) \cdot \bm{\epsilon}_\lambda(\bm{k}) = \delta_{\lambda \lambda'}\, , \quad \bm{\epsilon}^*_{\lambda}(\bm{k}) = \bm{\epsilon}_\lambda(-\bm{k}) \, ,
\ee
where $k \equiv |\bm{k}|$. From eq.~\eqref{PhotonEOM}, the EOM for the mode functions reads
\be
\label{PhotonEOMk}
\frac{\partial^2}{\partial \tau^2}\, A_\lambda(\tau, k) \, + \, k \left( k \, - \, \lambda \, \xi(\tau) \right) A_\lambda(\tau, k) \, = \, 0 \, ,
\ee
where
\be
\label{eq:xi}
\xi(\tau) \, \equiv \, - \cos^2 \theta_W  \frac{\partial_\tau h^2}{M^2} \, .
\ee
We solve eq.~\eqref{PhotonEOMk} using the evolution of the Higgs that follows from eq.~\eqref{HiggsEOMs}, starting from the end of inflation. Assuming $\xi \approx 0$ for $\tau\leq\tau_{\rm md}$, the initial conditions for the mode functions at $\tau=\tau_{\rm md}$ are given by\footnote{Note that if $\xi\neq0$ for $\tau\leq\tau_{\rm md}$, modes with momenta $k <|\xi|$ may already be excited above the vacuum. Using eq.~\eqref{Aic} for the initial conditions then corresponds to neglecting this contribution.}
\begin{subequations} 
\label{Aic}
\begin{eqnarray}
A_\lambda(\tau_{\rm md},k) &=& \dfrac{1}{\sqrt{2k}}   \\
 \pder{A_\lambda(\tau_{\rm md},k)}{\tau} &=& -i\sqrt{\dfrac{k}{2}} \, .
\end{eqnarray}
\end{subequations}

As follows from eq.~\eqref{PhotonEOMk}, modes with momenta $k < |\xi|$ and helicity $\lambda = \sign (\xi)$ have a tachyonic instability. This leads to the copious production of these modes. In addition, also modes with momenta $k > |\xi|$ can be produced via parametric resonance due to the time-dependence of the effective mass in eq.~\eqref{PhotonEOMk}. We will be interested in having an imbalance in the amplitudes of modes with positive and negative helicity that are produced. The net helicity density is given by
\be
\label{eq:Helicity}
\mathcal{H} \, \equiv \,  \lim_{V \rightarrow \infty} \frac{1}{V} \int_V d^3x \, \moy{\AB} =  \int d k \,  \frac{k^3}{2 \pi^2}\left(|A_+|^2-|A_-|^2\right) \, ,
\ee
where  the integral over $V$ averages the quantity over space and $\moy{...}$ denotes the expectation value of the operators.  
The magnetic field in eq.~\eqref{eq:Helicity} is given by $\bm{B}= \bm{\nabla} \times \bm{A}$.
Similarly, in the radiation gauge, the electric field is given by $\bm{E} = -\partial_\tau \bm{A}$. These are comoving quantities from which the physical electric and magnetic fields follow as $\hat{\bm{E}} = \bm{E}/a^2$ and $\hat{\bm{B}} = \bm{B}/a^2$. Hereafter we will denote physical quantities, as opposed to comoving ones, with a hat. The energy densities in the electric and magnetic field read
\begin{subequations} 
\label{EnergyDensityEB}
\begin{eqnarray}
\label{EnergyDensityE}
\rho_{E} \, & \equiv & \, \lim_{V \rightarrow \infty} \frac{1}{2 \, V} \int_V d^3x \, \moy{\bm{E}^2} \, = \int_0^{k_c} d k \,  \frac{k^2}{4 \pi^2}\left(|\partial_\tau A_+|^2 + |\partial_\tau A_-|^2\right) \\
\rho_{B} \, & \equiv & \, \lim_{V \rightarrow \infty} \frac{1}{2 \, V} \int_V d^3x \, \moy{\bm{B}^2} \, = \int_0^{k_c} d k \,  \frac{k^4}{4 \pi^2}\left(|A_+|^2 + |A_-|^2\right) .
\label{EnergyDensityB}
\end{eqnarray}
\end{subequations} 
The integral over momentum space diverges as $k^4$ for large momenta since $|\partial_\tau A_\lambda| = \sqrt{k/2}$ and $|A_\lambda| = 1/\sqrt{2 k}$ in the vacuum. We therefore impose a momentum cutoff $k_c$ which we choose as the largest momentum for which the corresponding modes are excited above the vacuum after production has shut off, $|A_\lambda| > c/\sqrt{2 k}$ with $c$ an $\mathcal{O}(1)$ constant. We have checked that our results do not depend sensitively on the value of this constant $c$. We will also need the correlation length of the magnetic field which can be estimated as \cite{Durrer:2013pga}
\be
\label{CorrelationLengthB}
\lambda_B \, = \, \frac{2 \pi}{\rho_B}\int d k \,  \frac{k^3}{4 \pi^2}\left(|A_+|^2 + |A_-|^2\right) .
\ee
Note that the helicity $\mathcal{H}$, the energy densities $\rho_{E,B}$ and the correlation length $\lambda_B$ are again comoving quantities. The corresponding physical quantities are given by $\hat{\mathcal{H}}= \mathcal{H} /a^3$, $\hat{\rho}_{E,B} = \rho_{E,B}/a^4$ and $\hat{\lambda}_B= a \lambda_B$.

Due to the chiral anomaly, the production of helicity is accompanied by the generation of fermionic asymmetries. This can be understood as arising from the Schwinger effect in a strong electromagnetic field \cite{Domcke:2018eki,Domcke:2019qmm}.
To discuss this, let us focus on the right-handed electron. The current associated with $U(1)$ rotations of the right-handed electron is anomalous and its divergence is given by
\be
\label{AnomalyEquation}
\partial_\mu J^\mu_{e_R} \, = \, -  \frac{\alpha}{4 \pi} \,  F_{\mu \nu} \tilde{F}^{\mu \nu} \, + \, \dots \, ,
\ee
where $\alpha=e^2/4 \pi$ is the fine-structure constant and the ellipsis denotes the contribution from the mass term of the electron. We typically have $\langle|\hat{\bm{E}}|\rangle \gg m_e^2$, where $m_e$ is the electron mass, in which case this contribution can be neglected \cite{Domcke:2019qmm}. Defining the charge 
\be
\label{AsymmetryDefinition}
q_{e_R} \, \equiv \, \lim_{V \rightarrow \infty} \frac{1}{V} \int_V d^3x \, \moy{J^0_{e_R}} 
\ee
corresponding to the asymmetry in the number densities of the right-handed electron and its antiparticle, the anomaly equation then gives
\be
\label{HelicityAnomalyEquation}
\partial_\tau q_{e_R} \, \simeq \, - \frac{\alpha}{2 \pi} \partial_\tau \mathcal{H} .
\ee
From this, we expect that the asymmetry
\be
\label{AsymmetryHelicityRelation}
q_{e_R} \, \simeq \, -  \frac{\alpha}{2 \pi} \mathcal{H} 
\ee
is generated together with the helicity. Similarly, asymmetries are also generated for the other charged fermions. We will discuss the potential backreaction of this process on gauge field production in sec.~\ref{sec:backreaction} and the subsequent evolution of the asymmetries in sec.~\ref{sec:HelicityEvolution}.

\subsection{Backreaction and finite-temperature effects}
\label{sec:backreaction}

We now comment on three issues that could affect the evolution of the Higgs and the production of photons. 
Firstly, the Higgs of course couples not only to the photon via the term in eq.~\eqref{eq:LEaction} but also to the other SM particles. The time evolution of the Higgs in particular leads to time-dependent masses for the latter which in turn results in the non-perturbative production of SM particles \cite{Enqvist:2013kaa}. The backreaction from this process can then affect the time evolution of the Higgs. Due to the Pauli exclusion principle, the production of fermions is suppressed. We can therefore focus on the massive gauge bosons, $W^\pm$ and the $Z$. For a Higgs potential and initial conditions similar to the ones that we will consider, it was found in \cite{Yang:2015ida} that the production of $W^\pm$ and $Z$ bosons via this process begins to affect the time evolution of the Higgs only after it has oscillated several times in its potential. This can be understood from the fact that the production mechanism is active only while the Higgs crosses zero and the SM particles become light. The Higgs therefore has to cross zero several times before enough energy is dumped into $W^\pm$ and $Z$ bosons to affect its evolution. This is in contrast to the production of photons via the coupling in eq.~\eqref{eq:LEaction} which is active also for large Higgs VEVs and is thus much more efficient. We will find that the majority of the helicity is produced during the first few oscillations and will therefore use eq.~\eqref{HiggsEOMs} for the Higgs, neglecting the backreaction from the production of $W^\pm$ and $Z$ bosons at early stages (while we check that the backreaction from photon production is small as discussed in sec.~\ref{sec:model}). Related to this, the Higgs can also decay perturbatively which leads to an additional damping term in eq.~\eqref{HiggsEOMs}. The Higgs decay rate is dominated by decays into $b$ quarks also for large Higgs VEVs and is estimated as  \cite{Enqvist:2013kaa}    
\be
\Gamma(h \rightarrow b \bar{b}) \, = \, \frac{3^{3/2} \sqrt{\lambda_h} \, y_b^2}{16 \pi} \, h \left( 1 - \frac{2 y_b^2}{3 \lambda_h}\right)^{3/2} ,
\ee
where $h$ is the Higgs VEV, $\lambda_h$ the (field-dependent) Higgs quartic coupling and $y_b \sim 10^{-2}$ the Yukawa coupling to the $b$ quark. 
For $h \ll 2 \cdot 10^4 H / \sqrt{\lambda_h}$, where $H$ is the Hubble rate, the damping term from perturbative decays is negligible compared to the Hubble-induced damping term in eq.~\eqref{HiggsEOMs}. This will always be fulfilled in the cases that we consider.  

Secondly, the presence of a strong electromagnetic field can lead to the pair production of charged particles via the Schwinger effect \cite{Schwinger:1951nm}. This proceeds via two channels.  One is asymmetric between left- and right-handed particles and leads to the asymmetries discussed in sec.~\ref{sec:GaugeFieldProduction}, while the other channel is symmetric \cite{Domcke:2018eki,Domcke:2019qmm}. The particles are subsequently accelerated in the electromagnetic field, resulting in a current \cite{Kobayashi:2014zza,Hayashinaka:2016qqn,Bavarsad:2017oyv,Domcke:2018eki,Domcke:2018gfr,Domcke:2019qmm}. This current drains energy from the electromagnetic field and thereby backreacts on its production. In our case, we have nonvanishing electric and magnetic fields and all charged particles are heavy while the Higgs rolls down its potential. For parallel and constant physical electric and magnetic fields, the resulting current was calculated in \cite{Domcke:2019qmm}. From eq.~\eqref{EnergyDensityEB}
the strengths of the electric and magnetic field are $\hat{E} \approx \sqrt{2 \hat{\rho}_E}$ and $\hat{B} \approx\sqrt{2 \hat{\rho}_B}$, respectively. The induced physical current $\hat{J}_{\rm ind}$ of a particle with mass $m$ and electric charge $e Q$ along the direction of parallel electric and magnetic fields is then determined by \cite{Domcke:2019qmm}
\be
\label{current}
e Q \, \frac{\partial}{\partial \tau}\left( a^3 \hat{J}_{\rm ind}\right) \, = \, a^4 \frac{(e |Q|)^3}{2 \pi^2} \hat{E} \hat{B} \coth\left(\frac{\pi \hat{B}}{\hat{E}} \right) \, \exp{-\frac{\pi m^2}{e |Q| \hat{E}}} \, .
\ee
The EOM for the gauge field including the contribution from the current leads to the conservation equation \cite{Domcke:2019qmm}
\be
\label{ConservationEquation}
a^{-1} \frac{\partial}{\partial \tau} \left(\hat{\rho}_E + \hat{\rho}_B \right) \, = \, - 4 H \left(\hat{\rho}_E + \hat{\rho}_B \right) \, + \, a^{-1} |\xi| \, \hat{E} \hat{B} \, - \, e Q \hat{E} \hat{J}_{\rm ind} \, .
\ee
The second-last term on the RHS arises from the production of gauge fields, while the last term is due to the backreaction from the current. Jumping ahead, we have applied eq.~\eqref{current} to the benchmark points studied in sec.~\ref{sec:result}. We find that the backreaction term in eq.~\eqref{ConservationEquation} becomes comparable to the production term toward the end of the time evolution, when the production mechanism becomes inefficient because the Higgs evolves too slowly. This means that the backreaction could become important. However, note that the physical electric and magnetic field in our case are far from constant but change quickly during the production process. The current may therefore differ from the one estimated from eq.~\eqref{current}. Furthermore, once the production mechanism switches off, the electric and magnetic field become free fields and orthogonal to each other. Subsequently, pair production no longer happens \cite{Schwinger:1951nm}. We leave a more thorough analysis of the backreaction from the current in our scenario for future work.

Finally, we comment on our assumption that the temperature vanishes while the magnetic helicity is produced and that reheating takes place only afterwards. The inflaton can produce particles nonperturbatively while it oscillates in its potential in a process known as preheating. This process can be strongly suppressed, however, if the inflaton couples dominantly to fermions instead of bosons or if its couplings are small. In this case, the reheating of the universe proceeds via perturbative decays of the inflaton.
Reheating is usually defined to happen when the Hubble rate has fallen sufficiently so that it equals the inflaton decay rate. 
The temperature at this time (assuming thermalization) is obtained from equating the two rates,
\be
\label{ReheatingTemperature}
\hat{T}_{\rm rh} \, = \, \left(\frac{90}{\pi^2 g_*} \right)^{1/4} \sqrt{\Gamma_{\rm inf} M_{\rm Pl}} \, ,
\ee
where $\Gamma_{\rm inf}$ is the inflaton decay rate, $M_{\rm Pl}$ the reduced Planck mass and $g_*$ the number of relativistic degrees of freedom at reheating. 
As has been pointed out in \cite{Chung:1998rq,Giudice:2000ex}, the temperature before reheating does not necessarily vanish though. Indeed for a constant decay rate, the inflaton already slowly decays during inflaton oscillations, leading to a plasma of particles. Whether and when this plasma thermalizes depends on the particles into which the inflaton decays and the strengths of their couplings. Assuming that it does thermalize, the temperature of the plasma quickly reaches a maximum after inflation \cite{Chung:1998rq,Giudice:2000ex}. For the case that the plasma consists of SM particles, this maximum temperature reads
\be
\label{MaximalTemperature}
\hat{T}_{\rm max} \, \approx \, 0.6 \,  \left(\frac{90}{\pi^2 g_*} \right)^{1/8}  (H_{\rm inf} M_{\rm Pl})^{1/4} \sqrt{\hat{T}_{\rm rh}} \, = \, 0.6 \,  \sqrt{\hat{T}_{\rm rh}^{\rm ins} \hat{T}_{\rm rh}}\, ,
\ee
where $\hat{T}_{\rm rh}^{\rm ins}$ is the temperature after instant reheating, i.e.~eq.~\eqref{ReheatingTemperature} for $\Gamma_{\rm inf}=H_{\rm inf}$. 
This maximal temperature is thus larger than the reheating temperature eq.~\eqref{ReheatingTemperature}. The temperature subsequently decreases and equals eq.~\eqref{ReheatingTemperature} at the time of reheating.
The presence of a plasma of SM particles (whether it is thermalized or not) before reheating could affect both the evolution of the Higgs and the production of magnetic helicity. Indeed the resulting thermal corrections in the Higgs potential could dominate due to the large temperature. 

Furthermore, charged particles in the plasma may backreact on and suppress the production of electromagnetic fields. We should therefore ensure that the energy density in charged particles is sufficiently small while the magnetic helicity is produced. Requiring that the interaction rate of the charged particles with the electromagnetic fields is smaller than the Hubble rate, one finds \cite{Fujita:2015iga}
\be
\label{PlasmaInteractionCondition}
\left(\frac{\alpha}{0.01} \right)^2 \left(\frac{\Gamma_{\chi \rightarrow \rm SM}}{2 \cdot 10^6 \, \text{GeV}} \right) \left(\frac{m_\chi}{10^{13}\, \text{GeV}} \right)^{-3} \, < \, 1,
\ee
where $\Gamma_{\chi \rightarrow \rm SM}$ is the inflaton decay rate into (charged) SM particles and $m_\chi$ is the inflaton mass. We will consider reheating temperatures $\hat{T}_{\rm rh} \sim 10^{14}\,$GeV which from eq.~\eqref{ReheatingTemperature} corresponds to inflaton decay rates $\Gamma_{\chi \rightarrow \rm SM} \sim 10^{10}\,$GeV. 
Even for inflaton masses $m_\chi \lesssim 10^{13}\, \text{GeV}$, the condition in eq.~\eqref{PlasmaInteractionCondition} can be fulfilled as follows: We will have Higgs VEVs $\gg H_{\rm inf}$ during the production process and the SM particles are thus very heavy. The inflaton decay rate $\Gamma_{\chi \rightarrow \rm SM}$ can be strongly suppressed if the inflaton is lighter than the SM particles that it dominantly couples to. The corresponding decay channels then become possible only at later times when the Higgs VEV has sufficiently decreased, leading $\Gamma_{\chi \rightarrow \rm SM}$ to increase rapidly. If $\Gamma_{\chi \rightarrow \rm SM}> H$ once this happens, the universe reheats quickly at this time. 
We then assume that this process takes place after the magnetic helicity has been produced. As an alternative option, the inflaton could couple to and decay into some new uncharged particles which couple to the SM sector sufficiently weakly that they reach equilibrium with the SM particles only after the magnetic helicity has been produced \cite{Fujita:2015iga}.

\section{Evolution of the helicity after reheating}
\label{sec:HelicityEvolution}

During the rolling and the oscillations of the Higgs in its potential, EW symmetry is broken except for the brief moments when the Higgs crosses zero. In sec.~\ref{sec:model}, we have therefore focused on the photon. $W^\pm$ and $Z$ bosons may also be produced whenever the Higgs crosses zero (either through the coupling in eq.~\eqref{eq:action} or through the gauge couplings) but we expect their contribution to be small and to decay quickly once the Higgs is again away from zero. In this section, we will consider the relevant quantities after reheating. 

\subsection{Helicity and fermionic asymmetries after reheating}
\label{sec:HelicityConversionReheating}

At reheating the EW symmetry is permanently restored by the thermal corrections (until the EW phase transition) and the helicity in photons is transformed into helicity in hypercharge gauge fields. Setting $A_{Y \mu} = (A_{Y 0},\bm{A}_Y)$ and using the radiation gauge, the helicity in hypercharge gauge fields $\mathcal{H}_Y$ is defined analogously to eq.~\eqref{eq:Helicity}. We then expect
\be
\label{HelicityRelation}
\mathcal{H}_Y = \cos^2\theta_W \, \mathcal{H}
\ee
after reheating, where $\mathcal{H}$ is the helicity in photons that was produced from the rolling and the oscillations of the Higgs. Note that part of the helicity is also converted into helicity in $W$ bosons but we expect this do be washed out quickly in the thermal plasma due to the thermal mass and self-interaction of the $W$ boson. 

As we have discussed in sec.~\ref{sec:backreaction}, we expect that asymmetries in the SM fermions are generated together with the helicity. In particular, the asymmetry in right-handed electrons is given in eq.~\eqref{AsymmetryHelicityRelation}. Ignoring additional contributions from the conversion of the helicity (which we expect to be small), using eq.~\eqref{HelicityRelation} this can be written as
\be
\label{AsymmetryHelicityRelation2}
q_{e_R} \, \simeq \, -  \frac{\alpha_Y}{2 \pi} \mathcal{H}_{Y} \, ,
\ee
where $\alpha_Y=g_Y^2/4 \pi$ is the hypercharge fine-structure constant. At temperatures below $\sim 10^{12}\,$GeV, EW sphalerons come into thermal equilibrium and lead to the rapid erasure of the asymmetries stored in left-handed fermions. Similarly, the asymmetries in right-handed fermions are driven to zero by sphalerons once right- and left-handed particles reach chemical equilibrium via their Yukawa couplings. Due to its small Yukawa coupling, the right-handed electron is the last species to reach chemical equilibrium, at temperatures $\sim 10^5 \,$GeV. Below this temperature, all the asymmetries are erased. We will see in sec.~\ref{sec:CPI}, though, that a process called {\it chiral plasma instability} can potentially convert the asymmetries into hypercharge helicity before they can be erased by sphalerons. Before addressing this, in the next section we will discuss the joint evolution of the hypercharge gauge fields with the thermal plasma. 

\subsection{Magnetohydrodynamics and Reynolds numbers}
\label{sec:MHD}

The hypercharge gauge fields interact with the thermal plasma of SM particles after reheating and set it into motion. This in turn backreacts on the hypercharge gauge fields. The combined system is described by Maxwell's equations and the Navier-Stokes equation (see \cite{Durrer:2013pga,Vachaspati:2020blt} for reviews). The relevant Maxwell's equations are
\be
\label{MaxwellsEquations}
\frac{\partial \bm{B}_Y}{\partial \tau} \, = \, - \bm{\nabla} \times \bm{E}_Y \qquad \quad \frac{\partial \bm{E}_Y}{\partial \tau} \, = \, \bm{\nabla} \times \bm{B}_Y \, - \, \bm{J}_Y\, ,
\ee
where the hypercharge electric and magnetic fields are given in terms of the hypercharge gauge field in radiation gauge by $\bm{E}_Y = -\partial_\tau \bm{A}_Y$ and $\bm{B}_Y = \bm{\nabla} \times \bm{A}_Y$, respectively. Furthermore, the current can be estimated from a generalized Ohm's law\footnote{Note that the $ \mu_{5}$-dependent term is only applicable for $ \mu_{5}/T \ll \alpha_Y$. This will always be fulfilled in our case. See \cite{Figueroa:2017hun} and references therein for other regimes.}
\be
\label{Current}
\bm{J}_Y \, = \, \sigma \, ( \bm{E}_Y \, + \, \bm{v} \times \bm{B}_Y ) \, + \, \frac{2 \alpha_Y}{\pi} \mu_{5} \bm{B}_Y\, ,
\ee
where $\bm{v}$ is the fluid velocity of the thermal plasma and $\sigma \simeq c_\sigma T / (\alpha_Y \log(\alpha_Y^{-1}))$ with $c_\sigma \approx 4.5$ is its conductivity \cite{Baym:1997gq,Arnold:2000dr}. Note that all quantities in eqs.~\eqref{MaxwellsEquations} and \eqref{Current} are comoving. In particular, $T=\hat{T} a$ is a comoving reference temperature, where $\hat{T}$ is the physical temperature. Up to changes in the number of degrees of freedom, $T$ stays constant during radiation domination. Furthermore, the last term in eq.~\eqref{Current} is due to the chiral magnetic effect \cite{Joyce:1997uy,Son:2004tq,Fukushima:2008xe} with
\be
\label{eq:mu5}
\mu_{5} \, = \, \sum_\alpha \epsilon_\alpha N_\alpha Y_{\alpha}^2 \mu_\alpha \, ,
\ee
where $\alpha$ runs over all SM species, with multiplicity $N_\alpha$ and hypercharge $Y_{\alpha}$, and $\epsilon_\alpha=\pm$ for right-/left-handed particles. Defining asymmetries $q_\alpha$ for the SM species in analogy with eq.~\eqref{AsymmetryDefinition}, $\mu_\alpha=6 q_\alpha /(N_\alpha T^2)$ for $\mu_\alpha \ll 1$ is the corresponding chemical potential that follows assuming kinetic equilibrium. 

For fluid velocities $|\bm{v}| \ll 1$, we can neglect the displacement current $\partial_\tau \bm{E}_Y$ in the Amp\`ere-Maxwell equation in eq.~\eqref{MaxwellsEquations}.\footnote{This can be seen as follows: Let us denote the characteristic electric and magnetic field and time and length scale of a gauge field configuration with $E_Y$, $B_Y$, $\tau_Y$ and $\lambda_{Y}$, respectively. We can estimate the terms in the Amp\`ere-Maxwell equation as $|\partial_\tau \bm{E}_Y| \sim E_Y /\tau_Y$ and $|\bm{\nabla} \times \bm{B}_Y|\sim B_Y /\lambda_{Y}$. Using the estimate $E_Y/B_Y \sim \lambda_{Y} /\tau_Y$ that follows from the Maxwell-Faraday equation, we then find $|\partial_\tau \bm{E}_Y| / |\bm{\nabla} \times \bm{B}_Y |\sim  (\lambda_{Y} / \tau_Y)^2 \sim |\bm{v}|^2 \ll 1$. } Combining this with eq.~\eqref{Current}, we can then solve for the hyperelectric field which gives
\be
\label{eq:HyperelectricField}
\bm{E}_Y \, = \, \frac{1}{\sigma} \bm{\nabla} \times \bm{B}_Y  - \frac{2 \alpha_Y}{\sigma \, \pi} \mu_{5} \bm{B}_Y - \, \bm{v} \times \bm{B}_Y \, .
\ee
Together with the Maxwell-Faraday equation in eq.~\eqref{MaxwellsEquations}, this yields the magnetohydrodynamics (MHD) equation for the hypermagnetic field
\be
\label{MHDeq1}
\frac{\partial}{\partial \tau} \bm{B}_Y \, = \, \frac{1}{\sigma} \bm{\nabla}^2\bm{B}_Y  + \bm{\nabla} \times ( \bm{v} \times \bm{B}_Y) +  \frac{2 \alpha_Y}{\pi} \frac{\mu_{5}}{\sigma} \bm{\nabla} \times \bm{B}_Y \, .
\ee
This is supplemented by the Navier-Stokes equation for the velocity field of an incompressible fluid interacting with the hypermagnetic field\footnote{Note that the viscous-damping term $\nu \bm{\nabla}^2\bm{v}$ is only present if the correlation length of the hypermagnetic field is larger than the mean free path of the particles in the plasma. This will always be fulfilled in our case. See \cite{BanerjeePhDThesis,Banerjee:2004df} for the damping term in the opposite regime.}
\be
\label{MHDeq2}
\frac{\partial}{\partial \tau} \bm{v} \, = \, \nu \, \bm{\nabla}^2\bm{v}  - (\bm{v} \cdot \bm{\nabla}) \, \bm{v}  + \frac{1}{\rho +p} (\bm{\nabla} \times \bm{B}_Y) \times \bm{B}_Y\, ,
\ee
where $\rho$ and $p$ are respectively  the energy and pressure density of the plasma. For radiation domination, we have $p=\rho/3$. Furthermore, $\nu \simeq c_\nu /(\alpha_Y^2 \log(\alpha_Y^{-1}) T )$ is the kinematic viscosity with $c_\nu \approx  0.01$ for temperatures above the EW scale \cite{Arnold:2000dr}. 

The MHD equations~\eqref{MHDeq1} and \eqref{MHDeq2} determine the coevolution of the hypermagnetic field and the fluid velocity of the thermal plasma. We are in particular interested in the evolution of the hypermagnetic helicity that is generated after inflation. By taking the time derivative of the hypermagnetic helicity defined analogously to eq.~\eqref{eq:Helicity} and using eq.~\eqref{eq:HyperelectricField}, we find
\be
\label{eq:HelicityEvolution}
\frac{\partial}{\partial \tau} \mathcal{H}_Y \, = \, \lim_{V \rightarrow \infty} \frac{-2}{V} \int_V d^3x \, \bm{E}_Y \cdot \bm{B}_Y\, = \, \lim_{V \rightarrow \infty} \frac{1}{V} \int_V d^3x \, \left( \frac{2}{\sigma} \bm{B}_Y \cdot \bm{\nabla}^2 \bm{A}_Y + \frac{4 \alpha_Y}{\pi}\frac{\mu_{5}}{\sigma} \bm{B}_Y^2 \right) \, .
\ee
Let us for the moment ignore the $\mu_5$-dependent terms in eqs.~\eqref{MHDeq1} and \eqref{eq:HelicityEvolution}. The hypermagnetic field can change due to magnetic diffusion and induction from the plasma motion, corresponding to the first and second term on the RHS of eq.~\eqref{MHDeq1}, respectively. Denoting the characteristic strength and correlation length of the magnetic field with $B_Y$ and $\lambda_{B_Y}$, respectively, and the typical velocity of the plasma at the length scale $\lambda_{B_Y}$ with $v$, we can estimate $\smash{|\bm{\nabla}^2\bm{B}_Y|/\sigma \sim B_Y/(\lambda_{B_Y}^2 \sigma)}$ and $\smash{ |\bm{\nabla} \times ( \bm{v} \times \bm{B}_Y)| \sim B_Y v/\lambda_{B_Y}}$. Induction then dominates over magnetic diffusion if the magnetic Reynolds number satisfies
\be
\label{eq:MagneticReynoldsNumber}
R_m \, \equiv \, \sigma \, v \, \lambda_{B_Y} \, \gtrsim \, 1\, .
\ee
Next notice that eq.~\eqref{eq:HelicityEvolution} does not depend on the plasma velocity which has dropped out. Induction from the plasma motion therefore does not lead to the decay of the helicity. We then expect that if the magnetic Reynolds number is larger than unity and the dynamics of the hypermagnetic field is dominated by the plasma motion, the helicity is preserved.

In order to discuss this in more detail, let us estimate the typical velocity $v$. To this end, we note that the last term on the RHS of eq.~\eqref{MHDeq2} acts as a source term that sets the plasma into motion. A steady velocity is obtained by balancing the first and second term with this source term. If the kinetic Reynolds number satisfies
\be
\label{eq:KineticReynoldsNumber}
R_e \, \equiv \, \frac{v \, \lambda_{B_Y}}{\nu} \, \gtrsim \, 1 \,,
\ee
the second term $\smash{|(\bm{v} \cdot \bm{\nabla}) \bm{v} |\sim v^2/\lambda_{B_Y}}$ dominates over the first term $\smash{\nu |\bm{\nabla}^2\bm{v}| \sim \nu \, v/\lambda_{B_Y}^2}$ and the typical velocity $v$ can be estimated as \cite{BanerjeePhDThesis,Banerjee:2004df}
\be
v \, \sim \, \frac{B_Y}{\sqrt{\rho}} \qquad \Longleftrightarrow \qquad \rho \, v^2 \, \sim \, B_Y^2 \, .
\ee
This corresponds to an equipartition between the kinetic energy in the plasma and the magnetic energy. If also the magnetic Reynolds number is larger than unity and the helicity is conserved, one finds the scaling relations (see e.g.~appendix B in \cite{Domcke:2019mnd} for a derivation)
\be
\label{eq:ScalingTurbulence}
B_Y \, \propto  \, \tau^{-\frac{1}{3}} \, , \qquad \lambda_{B_Y} \, \propto \,\tau^{\frac{2}{3}} \, , \qquad  v \, \propto \, \tau^{-\frac{1}{3}} \, .
\ee
From this, we see that the Reynolds numbers grow with time and thus remain larger than one. The conservation of the helicity, the equipartition of energy densities and the above scaling relations in this regime have been verified in numerical MHD simulations \cite{BanerjeePhDThesis,Banerjee:2004df,Kahniashvili:2012uj}. In particular, the conservation of the helicity can be understood as being due to an inverse cascade during which helicity is transferred from smaller to larger length scales. This is reflected in the growth of the characteristic length scale $\lambda_{B_Y}$ with time in eq.~\eqref{eq:ScalingTurbulence} and leads to the diffusion term in eq.~\eqref{eq:HelicityEvolution} being more and more suppressed over time. It thus never becomes important.

However, we find that in our scenario the kinetic Reynolds number is typically smaller than one. In this case, we can estimate the 
typical velocity $v$ by balancing the first and the last term on the RHS of eq.~\eqref{MHDeq2}. This gives \cite{BanerjeePhDThesis,Banerjee:2004df}
\be
\label{eq:EstimateVelocity}
v \, \sim \,  \frac{\lambda_{B_Y} B_Y^2}{\nu \rho} \qquad \Longleftrightarrow \qquad  \rho \, v^2 \, \sim \, R_e B_Y^2 
\ee
and the kinetic energy and velocity are thus smaller than for the case $R_e\gtrsim 1$. Using this in eq.~\eqref{eq:MagneticReynoldsNumber}, we expect that the helicity will be conserved at reheating if
\be
\label{eq:MagneticReynoldsNumber2}
R_m \, \sim \, \frac{\sigma \lambda_{B_Y}^2}{\nu} \frac{B_Y^2}{\rho}  \, \sim \,\left(\frac{40}{\pi^2 g_*}\right)^{1/2} \frac{c_\sigma \alpha_Y}{c_\nu} \frac{\rho_{B_Y} \lambda_{B_Y}^2 }{M_{\rm Pl} H_{\rm inf}} \left(\frac{\hat{T}_{\rm rh}}{\hat{T}_{\rm rh}^{\rm ins}} \right)^{2/3}\, \gtrsim\, 1 \, .
\ee
In the last step, we have used the estimate $B_Y^2 \approx 2 \rho_{B_Y}$, where $\rho_{B_Y}$ is the energy density calculated in analogy to eq.~\eqref{EnergyDensityB}. The correlation length of the hypermagnetic field $\lambda_{B_Y}$ can be calculated analogously to eq.~\eqref{CorrelationLengthB}. 
In the regime of small kinetic Reynolds number and assuming that the helicity is conserved, one finds the scaling relations (see again appendix B in \cite{Domcke:2019mnd})
\be
\label{eq:ScalingViscous}
B_Y \, \propto \, \tau^{-\frac{1}{2}} \, , \qquad \lambda_{B_Y} \, \propto \, \tau\, , \qquad  v \, \sim \, \text{const.}
\ee
As before, we see that the magnetic Reynolds number grows with time and thus stays larger than one. Correspondingly, we expect that the helicity is protected from diffusion and remains conserved if eq.~\eqref{eq:MagneticReynoldsNumber2} is fulfilled at reheating. The estimate for the velocity in eq.~\eqref{eq:EstimateVelocity} and the scaling relations in eq.~\eqref{eq:ScalingViscous} were previously derived in \cite{BanerjeePhDThesis,Banerjee:2004df} together with corresponding relations for the case that the velocity dissipates due to free streaming of particles instead of diffusion. A numerical MHD simulation was performed only for the latter case but the relations were verified with very good accuracy. We expect that the relations in our case of diffusion damping describe the evolution of the system similarly well but a numerical verification of this is clearly desirable. Note that also the kinetic Reynolds number grows with time too in the scaling regime eq.~\eqref{eq:ScalingViscous}. Eventually it may therefore become larger than unity and the quantities could subsequently scale as in eq.~\eqref{eq:ScalingTurbulence}. 

Alternatively, we can derive an upper bound on the magnetic Reynolds number without referring to the underlying MHD dynamics. Since the hypermagnetic field is the source of the velocity field, the kinetic energy of the latter is limited by the energy in the former, $\rho \, v^2 \lesssim B_Y^2 $. Plugging the velocity saturating this bound into eq.~\eqref{eq:MagneticReynoldsNumber}, we obtain the more conservative criterion
\be
\label{eq:MagneticReynoldsNumber3}
R_m^{\rm max} \, \sim \,  \sigma  \frac{\lambda_{B_Y} B_Y}{\sqrt{\rho}} \, \sim \, \left(\frac{40}{\pi^2 g_*}\right)^{1/4} \frac{c_\sigma }{\alpha_Y \log (\alpha_Y^{-1})} \frac{\lambda_{B_Y} \, \sqrt{\rho_{B_Y}}}{\sqrt{H_{\rm inf} M_{\rm Pl}}} \left(\frac{\hat{T}_{\rm rh}}{\hat{T}_{\rm rh}^{\rm ins}} \right)^{1/3}\, \gtrsim\, 1 
\ee
for the hypermagnetic fields to be able to survive until the EW phase transition. This is less stringent than eq.~\eqref{eq:MagneticReynoldsNumber2} (cf.~eq.~\eqref{eq:EstimateVelocity} for $R_e<1$).

\subsection{Chiral plasma instability}
\label{sec:CPI}

Let us next discuss the effect of the $\mu_5$-dependent terms in eqs.~\eqref{MHDeq1} and \eqref{eq:HelicityEvolution} which we have so far ignored. To this end, recall that asymmetries in the number densities of particles and their antiparticles are generated together with the helicity as discussed in sec.~\ref{sec:GaugeFieldProduction}. Since these asymmetries are related to the helicity via the chiral anomaly, they can be transformed back into helical gauge fields as we will now explain. Let us focus on the right-handed electron since it is the last species to come into chemical equilibrium, at temperatures $\sim 10^5\,$GeV, and its asymmetry thus survives the longest. The anomaly equation for the current corresponding to $U(1)$ rotations of the right-handed electron in the symmetric phase gives (cf.~eq.~\eqref{HelicityAnomalyEquation}) 
\be
\label{HelicityAnomalyEquation2}
\partial_\tau q_{e_R} \, \simeq \, - \frac{\alpha_Y}{2 \pi} \partial_\tau \mathcal{H}_Y \, ,
\ee
where we have assumed temperatures above $10^5 \,\text{GeV}$ and dropped the contribution from the Yukawa coupling.
From this, we see that if the asymmetry $q_{e_R}$ is driven to zero, a gauge field configuration with helicity $\mathcal{H}_Y \sim q_{e_R}/\alpha_Y$ is generated. Denoting the characteristic magnetic field of this configuration with $B_Y$ and its characteristic size with $\lambda_{Y}$, we can estimate its helicity in terms of these quantities using eq.~\eqref{eq:Helicity} as $ \mathcal{H}_Y \sim \lambda_{Y} B_Y^2$. The energy densities in right-handed electrons and the gauge field configuration are $\sim q_{e_R}^2/T^2$ and $\sim B_Y^2$, respectively. The gauge field configuration then has lower energy density than the equivalent asymmetry of right-handed electrons for \cite{Joyce:1997uy}
\be
\label{eq:CPIlength}
\lambda_{Y} \, \gtrsim \, \frac{T^2}{\alpha_Y q_{e_R}}  \, .
\ee
The formation of such an energetically favoured helicity configuration from an asymmetry is called the chiral plasma instability (CPI) \cite{Joyce:1997uy,Boyarsky:2011uy,Akamatsu:2013pjd,Hirono:2015rla,Yamamoto:2016xtu,Rogachevskii:2017uyc,Kamada:2018tcs}. 
In eq.~\eqref{eq:HelicityEvolution} for the helicity evolution, it arises from the $\mu_5$-dependent term. The fastest growing mode has a length scale saturating eq.~\eqref{eq:CPIlength}. Using that $\mu_5 \sim \mu_{e_R} \sim q_{e_R} /T^2$ in kinetic equilibrium if only right-handed electrons have an asymmetry and eqs.~\eqref{eq:HelicityEvolution} and \eqref{eq:CPIlength}, we can estimate the time scale of the CPI as
\be 
\label{eq:tauCPI}
\tau_{\rm CPI} \, \sim \, \frac{\sigma}{\alpha_Y^2 \mu_5^2}\, ,
\ee
where we have ignored all numerical prefactors. A more careful analysis of eqs.~\eqref{MHDeq1} and \eqref{eq:HelicityEvolution} in momentum space shows that eq.~\eqref{eq:tauCPI} also applies to the more general situation where several species have asymmetries and gives $\pi^2/2$ as a numerical prefactor \cite{Domcke:2019mnd}. 

Comparing eqs.~\eqref{AsymmetryHelicityRelation2} and \eqref{HelicityAnomalyEquation2}, we see that the CPI would convert the asymmetries into helicity which is approximately equal and opposite in sign to the helicity that is already present. This would thus strongly reduce the total helicity. 
In order to avoid this, we want to ensure that the CPI cannot occur before the electron Yukawa coupling reaches thermal equilibrium and all asymmetries are erased. To this end, we will require the time scale eq.~\eqref{eq:tauCPI} for the CPI to be sufficiently long that the universe has already cooled to temperatures below $\smash{10^5\,\text{GeV}}$ before it can happen. To determine the parameter $\mu_5$ in eq.~\eqref{eq:tauCPI}, let us consider temperatures somewhat above $\smash{10^5\,\text{GeV}}$ where all SM species except for the right-handed electron are in chemical equilibrium. Imposing constraints from sphalerons, Yukawa interactions and conserved quantities, the asymmetries and chemical potentials of all SM species can then be expressed in terms of those for the right-handed electron. Using eq.~\eqref{AsymmetryHelicityRelation2}, we get \cite{Domcke:2019mnd}
\be
\label{eq:mu5ElectronYukawa}
\mu_5 \, = \, - \alpha_Y \frac{2133}{481 \pi} \frac{\mathcal{H}_Y}{T^2} \, .
\ee

On the other hand, at temperatures below $10^5 \,$GeV, the electron Yukawa coupling is in thermal equilibrium and all asymmetries are erased. This gives $\mu_5=0$ and the CPI is no longer possible. Relating eq.~\eqref{eq:tauCPI} to the temperature of the universe at that time and using eq.~\eqref{eq:mu5ElectronYukawa}, we then demand that
\be
\label{TCPIcondition}
\hat{T}_{\rm CPI} \, \sim \, \frac{4 \, \alpha_Y^2 \mu_5^2}{\pi^2 \sigma H_{\rm inf}}\hat{T}_{\rm rh}^{1/3} (\hat{T}_{\rm rh}^{\rm ins})^{2/3}\, \sim \, \frac{g_* \alpha_Y^5 \log(\alpha_Y^{-1})}{\pi^2 c_\sigma}  \frac{\mathcal{H}_Y^2}{M_{\rm Pl}^2 H_{\rm inf}^3} \left( \frac{\hat{T}_{\rm rh}}{\hat{T}_{\rm rh}^{\rm ins}} \right)^2 \, \lesssim \, 10^5 \, \text{GeV}
\ee
to avoid the erasure of the hypermagnetic helicity by the CPI. Let us emphasize though that the derivation of this condition was necessarily approximate and that a numerical MHD simulation taking into account the asymmetries would be required to establish the condition for helicity survival in more detail.

\subsection{Conversion of the helicity at the electroweak phase transition}
\label{sec:HelicityConversionEWPT}

At temperatures around the EW scale, the Higgs obtains a VEV at $h\simeq 246\,$GeV and EW symmetry is broken. We assume no new physics which could affect this phase transition and it is thus a crossover. During the phase transition, the hypermagnetic helicity is converted back into magnetic helicity. The hypercharge gauge boson contributes to the $B+L$-anomaly, while the photon does not. 
The anomaly equation for the $B+L$-current then yields a relation similar to eq.~\eqref{HelicityAnomalyEquation} for the $B+L$-charge which in the unbroken phase depends on the hypermagnetic helicity but in the broken phase does not depend on the magnetic helicity. The conversion of the helicity at the EW phase transition therefore generates a $B+L$-asymmetry.  Sphalerons erase part of this asymmetry. As we will now explain though, since sphalerons switch off due to the broken EW symmetry at the same time as the asymmetry is being produced, a net asymmetry survives \cite{Kamada:2016cnb}.\footnote{We expect that a $B+L$-asymmetry is similarly produced at reheating where the inverse conversion of the helicity takes place. This asymmetry is subsequently completely erased by sphalerons and Yukawa interactions though as discussed in sec.~\ref{sec:HelicityConversionReheating}.}   
 
The conversion of the hypermagnetic fields into ordinary magnetic fields during the EW phase transition is governed by the EW angle $\theta_W$, i.e.~the angle of the $SO(2)$ rotation that diagonalizes the mass matrix for the gauge bosons  $A_{Y \mu}$ and $W_{\mu}^3$. The crucial point is that the EW angle changes smoothly since the thermal (magnetic) mass for (the transverse modes of) $W_{\mu}^3$ on the diagonal of the mass matrix\footnote{On the other hand, no magnetic mass arises for the hypercharge gauge boson or the photon \cite{Fradkin:1967qso,Kajantie:1996qd,DOnofrio:2015gop}.} initially dominates over the off-diagonal mass from the Higgs VEV which gradually develops during the crossover. The EW angle thus becomes a function of the temperature and smoothly changes from $\theta_W=0$ at high temperatures to $\theta_W=\arctan{g_Y/g_W}$ somewhat below the EW scale, where $g_Y$ and $g_W$ are the gauge couplings of respectively $U(1)_Y$ and $SU(2)_L$. This gives rise to a smooth source term for the $B+L$ asymmetry which is controlled by the changing EW angle. Above temperatures $\hat{T} \simeq  130\,$GeV \cite{DOnofrio:2014rug}, on the other hand, EW sphalerons are in thermal equilibrium and tend to erase the asymmetry. Including both contributions, the Boltzmann equation for the baryon-to-entropy ratio $\eta_B$ reads \cite{Kamada:2016cnb}
\be
\label{BoltzmannEquation}
\frac{d \eta_B}{d x}\, =\, -\frac{111}{34} \gamma_{W\rm sph} \, \eta_B \, + \, \frac{3}{16 \pi^2} (g_Y^2 + g_W^2) \sin(2 \theta_W) \frac{d \theta_W}{d x}  \, \frac{\mathcal{H}_Y}{s} \, ,
\ee
where $x = \hat{T}/H({\hat{T})}$ with $H({\hat{T}})$ being the Hubble rate at temperature $ \hat{T}$, $\mathcal{H}_Y$ is the hypermagnetic helicity that is initially present and $s=(2 \pi^2/45) g_* T^3$ is the comoving entropy density of the SM plasma. Furthermore, $\gamma_{W \rm sph}$ is the dimensionless transport coefficient for the EW sphaleron which for temperatures $\hat{T} <  161\,$GeV is found from lattice simulations as \cite{DOnofrio:2014rug}
\be
\gamma_{W \rm sph} \, \simeq \,  \exp{-147.7 + 107.9 \, \frac{\hat{T}}{130\, \text{GeV}}}  \, .
\ee 

The temperature-dependence of the EW angle $\theta_W$ has been determined analytically and from lattice simulations but is subject to significant uncertainties \cite{Kajantie:1996qd,DOnofrio:2015gop}. We follow \cite{Kamada:2016cnb,Jimenez:2017cdr} and model it with a smooth step function
\be
\label{SmoothStepFunction}
\cos^2 \theta_W \, = \, \frac{g_W^2}{g_Y^2+g_W^2} \, + \, \frac{1}{2}  \frac{g_Y^2}{g_Y^2+g_W^2}\left(1 + \tanh \left[\frac{\hat{T} - \hat{T}_{\rm step}}{ \Delta \hat{T}} \right] \right) 
\ee
which for $155\, \text{GeV}  \lesssim \hat{T}_{\rm step}  \lesssim 160\, \text{GeV} $ and $5\, \text{GeV} \lesssim \Delta \hat{T}  \lesssim 20\, \text{GeV} $ describes the analytical and lattice results for the temperature dependence reasonably well.

The Boltzmann equation \eqref{BoltzmannEquation} has been numerically solved in \cite{Kamada:2016cnb} and the baryon-to-entropy ratio $\eta_B$ was found to become frozen, i.e.~$\partial_\tau \eta_B=0$, at a temperature $\hat{T} \simeq 135 \,$GeV. As expected, this is close to the temperature $\hat{T} \simeq  130\,$GeV at which EW sphalerons freeze out. 
Setting the RHS of eq.~\eqref{BoltzmannEquation} to zero and solving for $\eta_B$, the observed baryon asymmetry of the universe is reproduced if 
\be
\label{eq:BAU}
\eta_B \, \simeq \, \frac{17}{1184\, \pi^2} \, (g_Y^2 + g_W^2)  \frac{\mathcal{H}_Y \hat{T}_{\rm rh}}{M_{\rm Pl}^2 H_{\rm inf}^2} \left[ \frac{f_{\theta_W}}{\gamma_{W \rm sph}} \frac{H({\hat{T}})}{\hat{T}} \right]_{\hat{T} =135\,\text{GeV}} \hspace{-.2cm}\simeq \, 9 \cdot 10^{-11} ,
\ee
where $f_{\theta_W}  \equiv -\sin (2 \theta_W) \, d\theta_W / d \log \hat{T}$. Varying $\hat{T}_{\rm step}$ and $\Delta \hat{T}$ in the ranges given below eq.~\eqref{SmoothStepFunction}, one finds $5.6 \cdot 10^{-4} < f_{\theta_W} < 0.32$ at $\hat{T} =135\,\text{GeV}$.

\subsection{Summary of constraints}

Let us summarize the constraints which we have derived. We need to generate gauge fields with helicity $\mathcal{H}_Y$, energy density $\rho_{B_Y}$ and correlation length $\lambda_{B_Y}$ which satisfy the condition on the magnetic Reynolds number eq.~\eqref{eq:MagneticReynoldsNumber2} (or at least eq.~\eqref{eq:MagneticReynoldsNumber3}) and the condition from the CPI eq.~\eqref{TCPIcondition} to survive until the EW phase transition and which fulfill eq.~\eqref{eq:BAU} to reproduce the observed baryon asymmetry of the universe. These conditions can be rewritten as\footnote{Note that the estimates for the magnetic Reynolds number, the CPI temperature and the baryon asymmetry depend on the gauge couplings which in turn depend on the RG scale $\mu$. For the magnetic Reynolds number, the hypercharge  gauge coupling should be evaluated near the reheating temperature. For concreteness, we have chosen the renormalization scale $\mu =10^{14} \, \text{GeV}$ in the estimates. Similarly, we have set $\mu =10^5 \, \text{GeV}$ in the estimate for the CPI temperature. In both cases, the dependence on the precise value of $\mu$ is quite weak.
}
\be
\label{ConditionsEstimateBaryonAsymmetry}
\eta_B  \, \simeq \,   2 \cdot 10^{-12} \, f_{\theta_W}  \frac{\mathcal{H}_Y}{H_{\rm inf}^3} \left( \frac{H_{\rm inf}}{10^{13} \, \text{GeV}} \right)^{3/2} \left(\frac{\hat{T}_{\rm rh}}{\hat{T}_{\rm rh}^{\rm ins}} \right) \, \, \simeq \, 9 \cdot 10^{-11} \, ,
\ee
where $\smash{5.6 \cdot 10^{-4} < f_{\theta_W} < 0.32}$ is evaluated at $\hat{T} =135\,\text{GeV}$, and
\bse 
\label{ConditionsEstimateTCPI}
\hat{T}_{\rm CPI} & \, \sim \,  & 2 \cdot 10^{-7} \, \text{GeV} \, \frac{\mathcal{H}_Y^2}{H_{\rm inf}^6} \, \left( \frac{H_{\rm inf}}{10^{13} \, \text{GeV}} \right)^3 \left(\frac{\hat{T}_{\rm rh}}{\hat{T}_{\rm rh}^{\rm ins}} \right)^2  \, \, \, \, \lesssim \, 10^5 \, \text{GeV} \, ,\\
\label{ConditionsEstimateReynoldsNumber}
R_m & \, \sim \,  &4 \cdot 10^{-6} \, \frac{\rho_{B_Y} \lambda_{B_Y}^2}{H_{\rm inf}^2} \, \left( \frac{H_{\rm inf}}{10^{13} \, \text{GeV}} \right)\left(\frac{\hat{T}_{\rm rh}}{\hat{T}_{\rm rh}^{\rm ins}} \right)^{2/3}   \,  \gtrsim \,  1\, ,
\ese
or alternatively
\be
\label{ConditionsEstimateReynoldsNumberMax}
R_m^{\rm max} \, \sim \,  8 \cdot 10^{-2}  \frac{\lambda_{B_Y}  \sqrt{\rho_{B_Y}}}{H_{\rm inf}} \left(\frac{H_{\rm inf}}{10^{13} \, \text{GeV}} \right)^{1/2}\left(\frac{\hat{T}_{\rm rh}}{\hat{T}_{\rm rh}^{\rm ins}} \right)^{1/3}\, \gtrsim\, 1 \, .
\ee
Note that the dimensionless combinations $\mathcal{H}_Y/H_{\rm inf}^3$, $\rho_{B_Y} /H_{\rm inf}^4$ and $ \lambda_{B_Y}H_{\rm inf}$ in the conditions are independent of $H_{\rm inf}$ if we keep $M^2/H_{\rm inf}^2$ and $h(\tau_{\rm md})/H_{\rm inf}$ fixed, where $h(\tau_{\rm md})$ is the initial Higgs VEV. To see this, rescale the Higgs, photon, conformal time and momentum by powers of $H_{\rm inf}$ to make them dimensionless. The solutions to the EOMs of the Higgs and the photon, eqs.~\eqref{HiggsEOMs} and \eqref{PhotonEOMk}, after this rescaling depend on $H_{\rm inf}$ only via the dimensionless ratios $M^2/H_{\rm inf}^2$, cf.~eq.~\eqref{eq:xi}, and $h(\tau_{\rm md})/H_{\rm inf}$. This is therefore also the case for the rescaled helicity, magnetic energy density and correlation length which are derived from these solutions. 

The constraints all depend on the reheating temperature $\hat{T}_{\rm rh}$ and we need to obtain a range for $\hat{T}_{\rm rh}$ for which they are all simultaneously fulfilled. In the next section, we will discuss a model which achieves that. 


\section{Higgs potential and numerical results}
\label{sec:result}

We are now in a position to study examples for the evolution of the Higgs after the end of inflation and to see whether this can generate the baryon asymmetry of the universe.  We have seen in the last section that three conditions need to be fulfilled. These are eq.~\eqref{ConditionsEstimateTCPI} from the CPI and eq.~\eqref{ConditionsEstimateReynoldsNumber} (or at least eq.~\eqref{ConditionsEstimateReynoldsNumberMax}) on the magnetic Reynolds number to ensure that the helicity survives until the EW phase transition. The observed baryon asymmetry is then reproduced if eq.~\eqref{ConditionsEstimateBaryonAsymmetry} is satisfied. 
For fixed $M^2/H_{\rm inf}^2$ and $h(\tau_{\rm md})/H_{\rm inf}$, the baryon-to-entropy ratio, the magnetic Reynolds number and the CPI temperature all grow with the inflation scale. In order to increase the first two parameters, we set $H_{\rm inf} = 10^{13}\,$GeV for the Hubble rate at the end of inflation. 
This leaves some margin to satisfy the upper limit on the Hubble rate from bounds on the tensor-to-scalar ratio (i.e.~roughly 60 e-folds before the end of inflation), $H_{\rm inf} < 6.1 \cdot 10^{13}\,$GeV \cite{Akrami:2018odb}.

Let us first  assume no new physics beyond the SM (except for the inflaton). During inflation, the Higgs is driven to large VEVs due to quantum fluctuations in de Sitter space. Depending on the precise values of SM parameters (in particular the top Yukawa coupling), the Higgs quartic coupling runs to negative values at energies as low as $\sim 10^{9}\, \text{GeV}$ \cite{Degrassi:2012ry}. For inflation scales larger than this instability scale, the Higgs is driven into the resulting AdS minimum, with catastrophic consequences \cite{Espinosa:2015qea}. 

We will therefore assume new physics which causes the Higgs quartic coupling to either not run negative or only at much larger energy scales than in the SM.\footnote{If the top mass is near the lower value of its measured range at $3\sigma$, the Higgs quartic coupling can remain positive up to the Planck scale \cite{Degrassi:2012ry}. In this case, we would not need new physics to stabilize the Higgs potential. Since it is experimentally disfavoured, however, we will not investigate this case further.} 
Quantum fluctuations during inflation are then expected to drive the average Higgs VEV to values $\sim \lambda_h^{-1/4} H_{\rm inf}$, where $\lambda_h$ is the Higgs quartic coupling (see e.g.~\cite{Enqvist:2013kaa,Espinosa:2015qea}). Starting from this VEV, the Higgs rolls down its potential toward the minimum once inflation ends. Assuming the coupling in eq.~\eqref{eq:action}, this leads to the production of electromagnetic fields. Let us consider the constraint on the magnetic Reynolds number to ensure their survival until the EW phase transition. The energy density of the hypermagnetic fields is bounded by the initial potential energy density of the Higgs, $\rho_{B_Y} \lesssim H_{\rm inf}^4$. We then see from eq.~\eqref{ConditionsEstimateReynoldsNumber} that even for instant reheating,  
$\hat{T}_{\rm rh} =\hat{T}_{\rm rh}^{\rm ins}$, the correlation length of the hypermagnetic fields has to satisfy $ \lambda_{B_Y} \gtrsim 500 \, H_{\rm inf}^{-1}$ in order to obtain a magnetic Reynolds number above one. For consistency though, we need $\hat{T}_{\rm rh} < \hat{T}_{\rm rh}^{\rm ins}$ since we have assumed a phase of matter domination and negligible temperature in the SM sector while the Higgs rolls down its potential and produces the magnetic fields. This makes the bound on the correlation length even stronger. We find, however, that typically $\lambda_{B_Y} \lesssim H_{\rm inf}^{-1}$ and this bound cannot be satisfied with the above initial conditions for the Higgs. Instead we will arrange for the Higgs to have a VEV $\gg H_{\rm inf}$ directly after inflation. This raises the energy density in the Higgs which can be converted into electromagnetic fields and thereby allows us to satisfy the constraint from the magnetic Reynolds number.

\subsection{Higgs potential with coupling to the Ricci scalar}

\label{sec:HiggsPotential}

There are of course many ways to affect the running of the Higgs quartic coupling. For definiteness, we focus on the simple possibility of a real scalar $\phi$ which couples to the Higgs doublet. Imposing the $\mathbb{Z}_2$ symmetry $\phi \rightarrow -\phi$, the most general renormalizable potential for $\phi$ and the Higgs doublet reads 
\be
V\, \supset \, -m_h^2 |\Phi|^2  \, + \, \lambda_h |\Phi|^4  \, + \, \frac{1}{2} m_\phi^2 \phi^2   \, + \, \lambda_\phi \phi^4 \, + \, \lambda_{\phi h} \phi^2 |\Phi|^2 \, .
\ee

We next include the coupling
\be
\label{eq:RicciScalarCoupling}
V\, \supset \, \xi_R R \, |\Phi|^2 \,,
\ee
where $R$ is the Ricci scalar. The background value of $R$ leads to a time-dependent contribution to the Higgs mass. 
As we review in appendix \ref{appendix:HiggsPotentialInflation}, $R=-6 [2-\epsilon(\tau)] H^2(\tau)$ during inflation, where $H(\tau)$ is the Hubble rate and $\epsilon(\tau)$ the slow-roll parameter.

\begin{figure}
\centering
\begin{subfigure}{.5\textwidth}
  \centering
 \includegraphics[width=6.8cm]{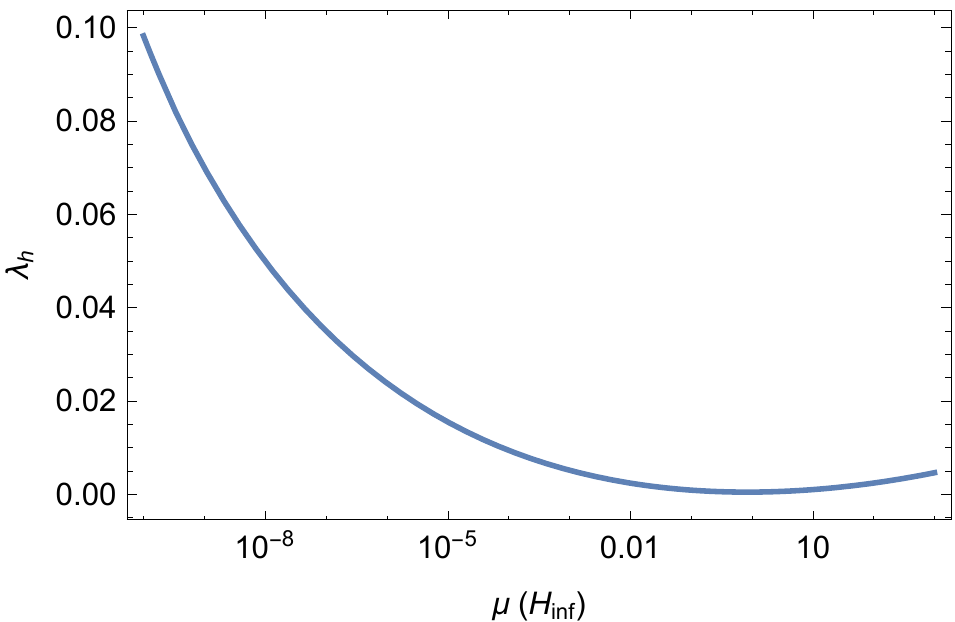}
\end{subfigure}%
\begin{subfigure}{.5\textwidth}
\centering
\includegraphics[width=7.5cm]{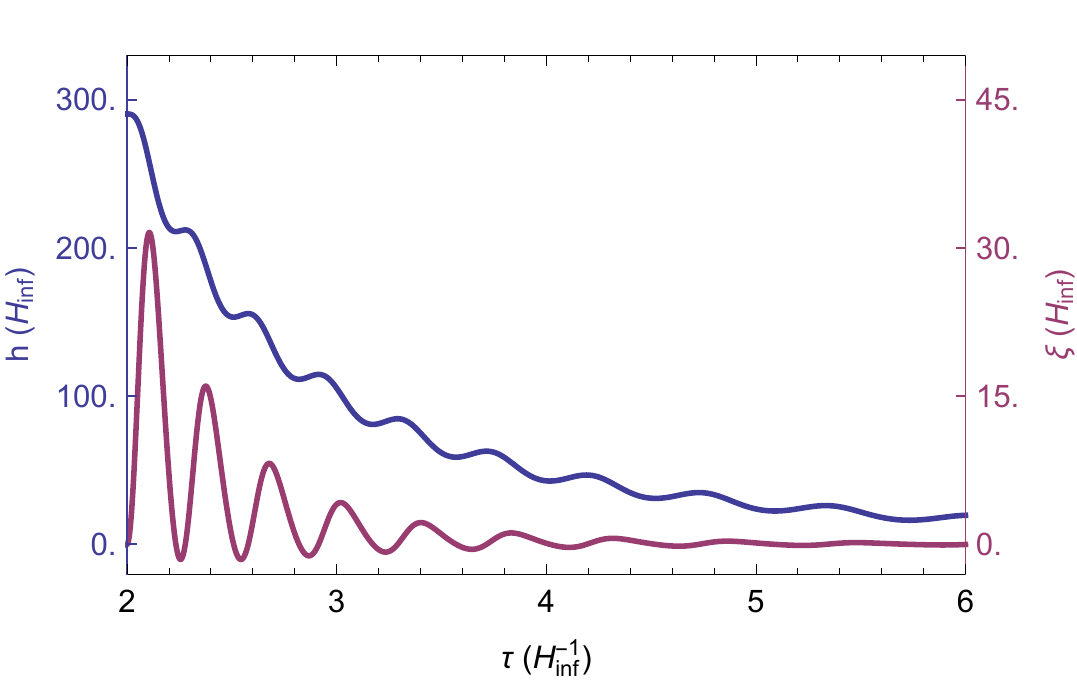}
\end{subfigure}
\begin{subfigure}{.5\textwidth}
  \centering
  \includegraphics[width=7.5cm]{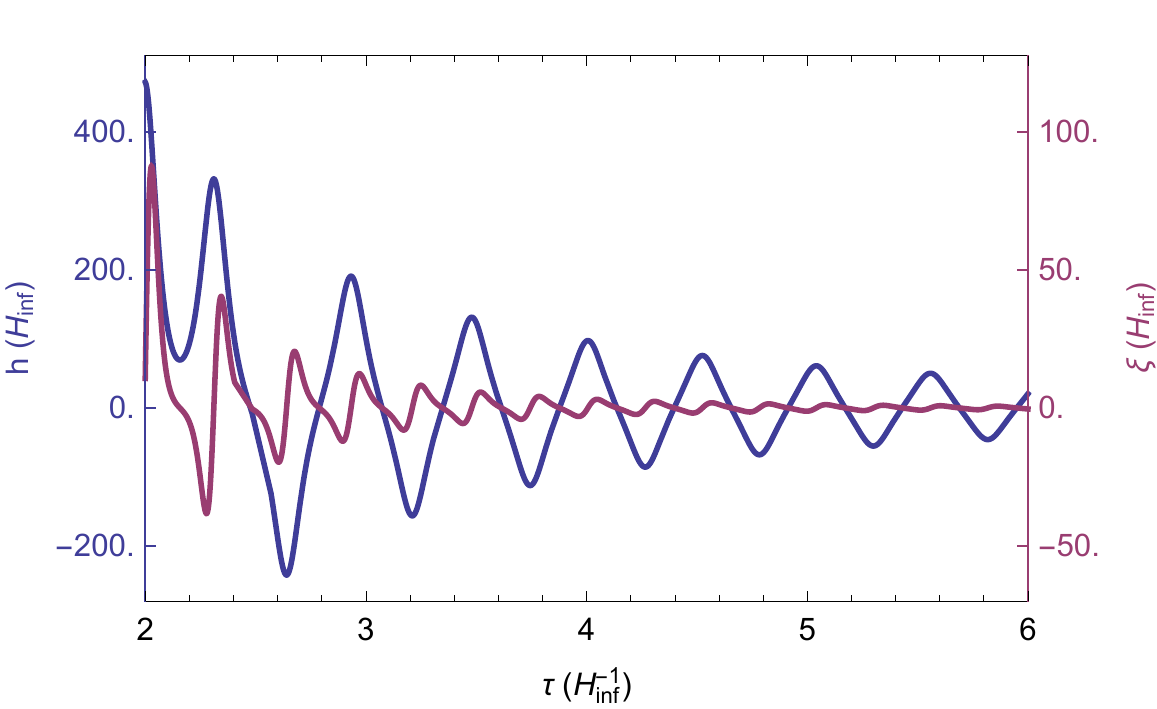}
\end{subfigure}%
\begin{subfigure}{.5\textwidth}
  \centering
  \includegraphics[width=7.5cm]{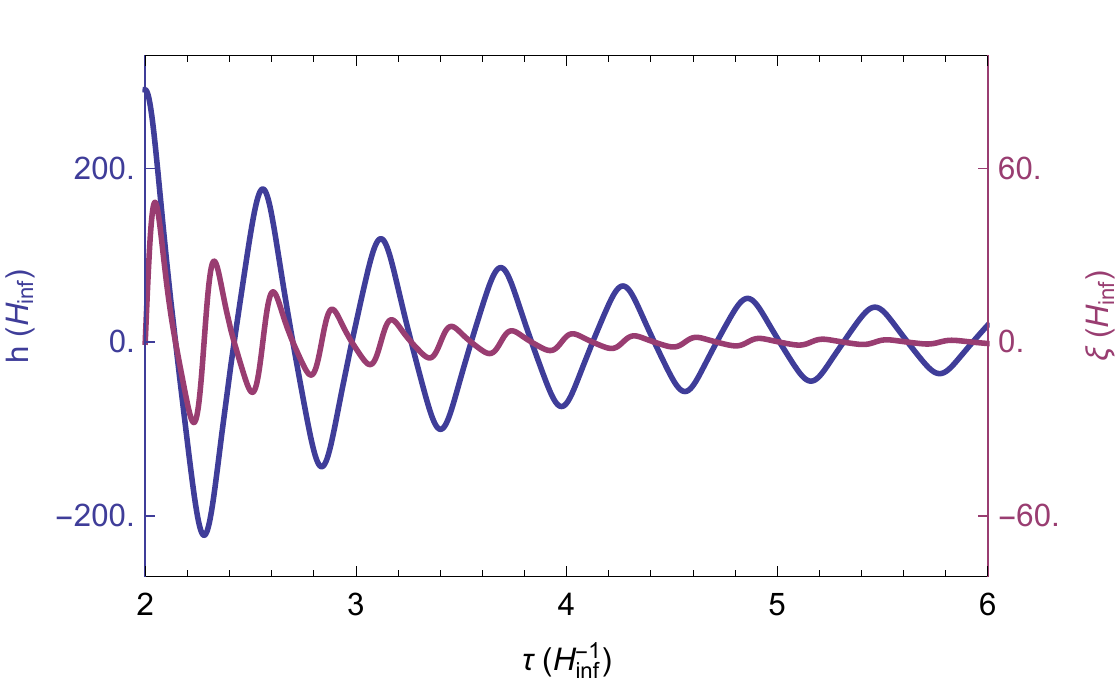}
  \end{subfigure}
\caption{\it Upper left panel: The running of the Higgs quartic coupling for $m_\phi = 4 \,$TeV, $\lambda_{\phi h}=0.2$ and $\lambda_{\phi }=0.01$ at the EW scale. The scale $\mu$ is shown in units of $H_{\rm inf}= 10^{13}\,$GeV. Upper right panel and lower panels: The time evolution of the Higgs $h$ (blue) and the instability parameter $\xi$ (purple) in the EOM of the gauge field \eqref{PhotonEOMk}. The upper right panel is for point 1 in table~\ref{table:BenchmarkPointsDefinition}, the lower left panel for point 2 and the lower right panel for point 3.}
\label{fig:xih}
\end{figure}

We choose $\smash{m_\phi^2 , \lambda_{\phi }, \lambda_{\phi h} > 0}$ (and the couplings remain positive after taking their running into account). This ensures that the minimum for $\phi$ is always at the origin. We will arrange for the Higgs to have a VEV 
$\gg H$ during inflation. This yields an effective mass $\gg H$ for $\phi$. De Sitter fluctuations of $\phi$ are then suppressed and it is anchored at the origin during inflation \cite{Espinosa:2015qea}. It is therefore a spectator field which enters the dynamics only via its loop corrections to the Higgs. We can take the loop corrections into account by calculating the RG-improved effective potential. To a good approximation this amounts to replacing the parameters in the tree-level potential by running parameters evaluated at the scale $\mu = h$. Setting $\phi = 0$, the effective potential for the Higgs during inflation reads\footnote{In regions of the potential with $h <  m_{\phi}$, one has to appropriately decouple the scalar in the RGEs. Since we choose $m_\phi \sim \text{few TeV}$, while $h \gg H_{\rm inf} = 10^{13} \,$GeV during most of the evolution that is relevant for us, we can ignore this in practice.}
\be
V_{\rm eff} \, \simeq \, - 3 [2-\epsilon(\tau)] \, \xi_R  H^2(\tau) \, h^2 \, + \, \frac{\lambda_h(h)}{4} \, h^4  .
\ee
We have neglected the Higgs mass parameter $m_h^2$ since it is very small compared to the mass induced by the Ricci scalar. The coupling $\xi_R$ runs very slowly and changes by at most $20\%$ between the Planck scale and the EW scale (see e.g.~\cite{Herranen:2014cua}). We therefore neglect its running and use a fixed value. For the Higgs quartic coupling, we use the RGEs at one-loop for the SM plus the singlet which we summarize in appendix~\ref{appendix:RGEs}. We choose values for the singlet mass $m_\phi$ and the couplings $\lambda_\phi$ and $\lambda_{\phi h}$ such that the Higgs quartic coupling remains always positive when running from the EW scale up. An example is shown in the upper left panel of fig.~\ref{fig:xih}.

For $\xi_R>0$, the Higgs potential has a minimum away from the origin which is induced by the tachyonic mass term due to the Ricci scalar. In most inflation models, $\epsilon \sim 1 $ toward the end of inflation so $\epsilon(\tau_{\rm inf})=1$ can be taken to define the conformal time $\tau_{\rm inf}$ when inflation ends. This gives $R=-6 H_{\rm inf}^2$ at the end of inflation, where we have used that by definition $H_{\rm inf} = H(\tau_{\rm inf})$. In hybrid inflation models, on the other hand, the inflaton stays in the slow-roll regime and the end of inflation is instead triggered by a second, waterfall field. In this case, we expect $\epsilon \ll 1$ and $R=-12 H_{\rm inf}^2$ at the end of inflation. Using this, the minimum of the Higgs potential at the end of inflation is determined by
\be
\label{HiggsPotentialMinimum}
h_{\rm min} \, \simeq \, \sqrt{\frac{c_R \, \xi_R}{\lambda_h + \beta_{\lambda_h}/(64\pi^2)}} \, H_{\rm inf}  \, ,
\ee
where $\beta_{\lambda_h}$ is the $\beta$-function of $\lambda_h$ given in eq.~\eqref{betafunctions} in appendix~\ref{appendix:RGEs} and $c_R=12$ for hybrid inflation or $c_R=6$ otherwise. All couplings on the RHS of eq.~\eqref{HiggsPotentialMinimum} have to be evaluated at the scale $\mu = h_{\rm min}$. We assume that the Higgs sits in this minimum at the end of inflation, $\smash{h(\tau_{\rm inf})=h_{\rm min}}$. For $\smash{h_{\rm min} \gg H_{\rm inf}}$, the resulting Higgs mass is much larger than $H_{\rm inf}$ and its fluctuations around the minimum are suppressed. This ensures that our baryogenesis mechanism does not lead to unacceptably large isocurvature perturbations \cite{Kusenko:2014lra,Pearce:2015nga,Yang:2015ida}. Furthermore, as discussed in appendix~\ref{appendix:HiggsPotentialInflation}, we find that 
$h'(\tau_{\rm inf})\simeq 0$ for hydrid inflation and $h'(\tau_{\rm inf}) \simeq - h_{\rm min}  H_{\rm inf}$ otherwise. 

Except for hybrid inflation, the velocity of the Higgs at the end of inflation is thus sizeable. This could lead to an additional contribution to photon production from the last e-folds of inflation which would enhance the helicity. However, the velocity of the Higgs during inflation is very suppressed and grows with the slow-roll parameter $\epsilon$ only toward the end (see eq.~\eqref{eq:velocity} in appendix \ref{appendix:HiggsPotentialInflation}). The size of this additional contribution therefore depends on how fast the slow-roll parameter increases from $\epsilon \ll 1$ to $\epsilon =1$ at the end of inflation and thus on the inflaton potential. In the following, we ignore the contribution to photon production during inflation and leave it to future work.

\begin{table}[t]
\centering 
\renewcommand{\arraystretch}{1.3}
\begin{tabular}{| c | c | c | c | c |} 
\hline 
 point & init.~cond. & $\xi_R$ &  $M/\text{GeV}$ & $h(\tau_{\rm md})/ H_{\rm inf} $\\ 
\hline 
1  & A  & 100 & $0.9 \cdot 10^{15}$ & $290.7$ \\
2  & B & 150 &  $1.8 \cdot 10^{15} $ & $471.7$\\
3  & C & 25 & $1.3 \cdot 10^{15}$ & $290.7$ \\
\hline 
\end{tabular}
\caption{\it Definition of three benchmark points,  setting $m_\phi = 4 \,$TeV, $\lambda_{\phi h}=0.2$ and $\lambda_{\phi }=0.01$ at the EW scale and $H_{\rm inf}= 10^{13}\,$GeV. For point 1 and 2, we have fixed $\hat{T}_{\rm rh} = 0.9 \cdot 10^{14}\, \text{GeV}$ and for point 3, we have chosen $\hat{T}_{\rm rh} = 1.5 \cdot 10^{14}\, \text{GeV}$. This corresponds to reheating happening at $\tau \approx 19 \, H_{\rm inf}^{-1}$ and $\tau \approx 14 \, H_{\rm inf}^{-1}$, respectively.  } 
\label{table:BenchmarkPointsDefinition} 
\end{table}

During the subsequent phase of inflaton oscillations, the Ricci scalar briefly changes its sign each time the inflaton crosses the origin of its potential (see e.g.~\cite{Figueroa:2017slm}). If the oscillation frequency of the inflaton is larger than the Hubble rate at the end of inflation, it oscillates many times during one Hubble time. We can then use the averaged value $R=-3 H^2(\tau)$, where $H(\tau)= H_{\rm inf} (\tau_{\rm md}/\tau)^3$ is the Hubble rate during matter domination.\footnote{Notice that for simplicity we identify the Hubble rates at the end of inflation and the onset of matter domination, $H(\tau_{\rm md}) = H(\tau_{\rm inf})= H_{\rm inf}$. Depending on how much time is spent until the Ricci scalar evolves as in matter domination, the Hubble rate at the latter time may be somewhat smaller.} We will focus on this case. Note that in the opposite case of slow inflaton oscillations, the sign changes of the Ricci scalar could induce additional motion in the Higgs, potentially increasing gauge field production. This case is left for future work. The effective potential for the Higgs during matter domination then reads
\be
\label{HiggsPotentialMatterDomination}
V_{\rm eff} \, \simeq \, -\frac{3}{2} \, \xi_R \, H_{\rm inf}^2 \left(\frac{\tau_{\rm md}}{\tau}\right)^6 h^2 \, + \, \frac{\lambda_h(h)}{4} \, h^4  \, .
\ee
The minimum of this potential at the onset of matter domination follows from eq.~\eqref{HiggsPotentialMinimum} with $c_R=3$. At later times, the minimum scales approximately like $(\tau_{\rm md}/\tau)^3$. 
\begin{figure}
\label{QuarticCoupling}
\end{figure}

Notice that the minima of the Higgs potential at the end of inflation and the onset of matter domination differ. This happens since $R= -12 H_{\rm inf}^2$ or $R= -6 H_{\rm inf}^2$ at the former time, while $R= -3 H_{\rm inf}^2$ at the latter time.
From this, we expect three extremal possibilities for the initial conditions of the Higgs at the onset of matter domination: If the Ricci scalar changes slowly between its values at the end of inflation and at the onset of matter domination, the Higgs can track the minimum of its potential and $h(\tau_{\rm md})=h_{\rm min}$ with $h_{\rm min}$ determined by eq.~\eqref{HiggsPotentialMinimum} for $c_R=3$. We then expect that the initial velocity of the Higgs is small and set $h'(\tau_{\rm md})=0$. If the Ricci scalar changes quickly, on the other hand, the Higgs is not able to follow the minimum and its initial value $h(\tau_{\rm md})=h_{\rm min}$ is determined by eq.~\eqref{HiggsPotentialMinimum} with $c_R=12$ for hydrid inflation or $c_R=6$ otherwise. The initial velocity then is $h'(\tau_{\rm md})\simeq 0$ or $h'(\tau_{\rm md})\simeq - h_{\rm min}  H_{\rm inf}$, respectively. To summarize we consider the three different initial conditions
\begin{enumerate}
\item[A:] $\; h(\tau_{\rm md})=h_{\rm min}\,$ for $\,c_R=3\,$ and $\,h'(\tau_{\rm md})=0$\,,
\item[B:] $\; h(\tau_{\rm md})=h_{\rm min}\,$ for $\,c_R=6\,$ and $\,h'(\tau_{\rm md})=- h_{\rm min} H_{\rm inf}$\,,
\item[C:] $\; h(\tau_{\rm md})=h_{\rm min}\,$ for $\,c_R=12\,$ and $\,h'(\tau_{\rm md})=0$\,.
\end{enumerate}
These initial conditions are clearly idealizations but we expect that they cover the range of possibilities relevant for baryogenesis. In particular, keeping the parameters which determine the Higgs potential eq.~\eqref{HiggsPotentialMatterDomination} fixed, initial condition C allows for significantly stronger gauge field production than A in the regime where the backreaction can be neglected. Initial condition B, on the other hand, lies in the middle.

\begin{figure}
\centering
\begin{subfigure}{.5\textwidth}
  \centering
  \includegraphics[width=7cm]{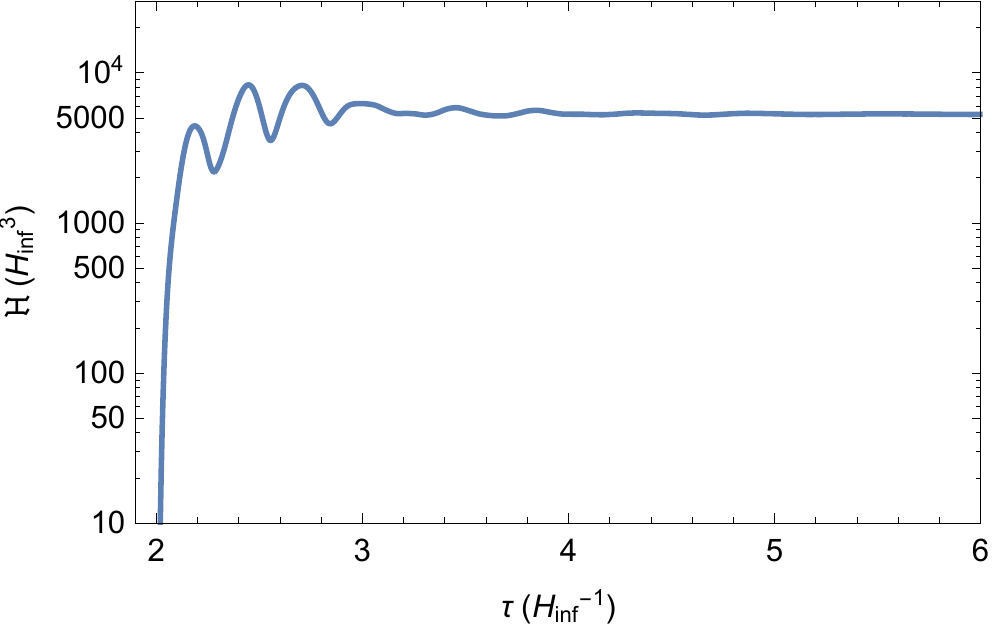}
\end{subfigure}%
\begin{subfigure}{.5\textwidth}
  \centering
  \includegraphics[width=7cm]{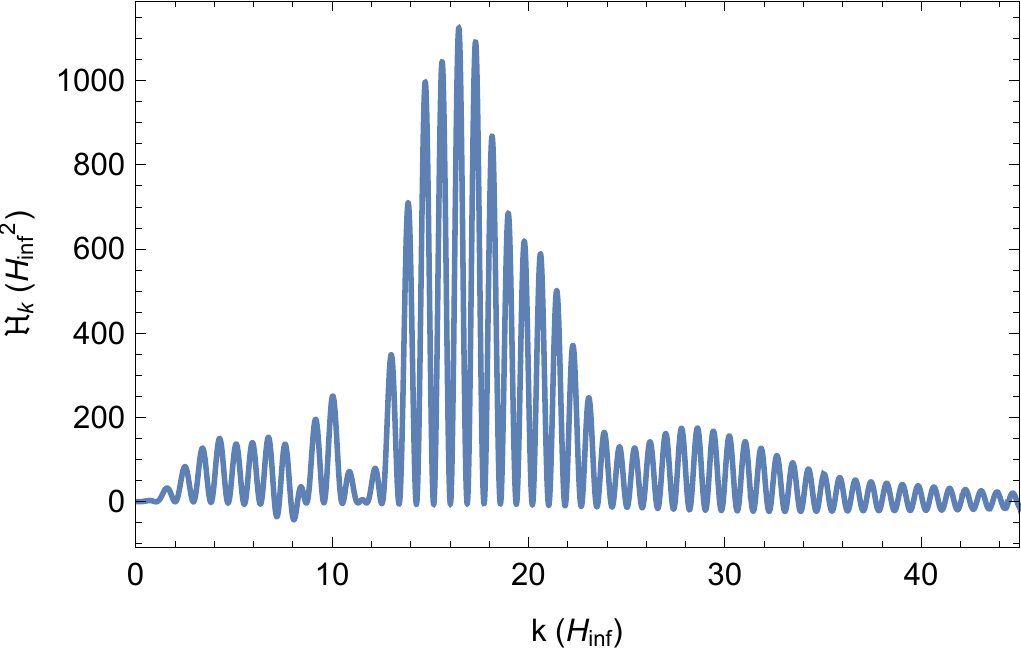}
  \end{subfigure}\\[1ex]
  \begin{subfigure}{.5\textwidth}
  \centering
  \includegraphics[width=7cm]{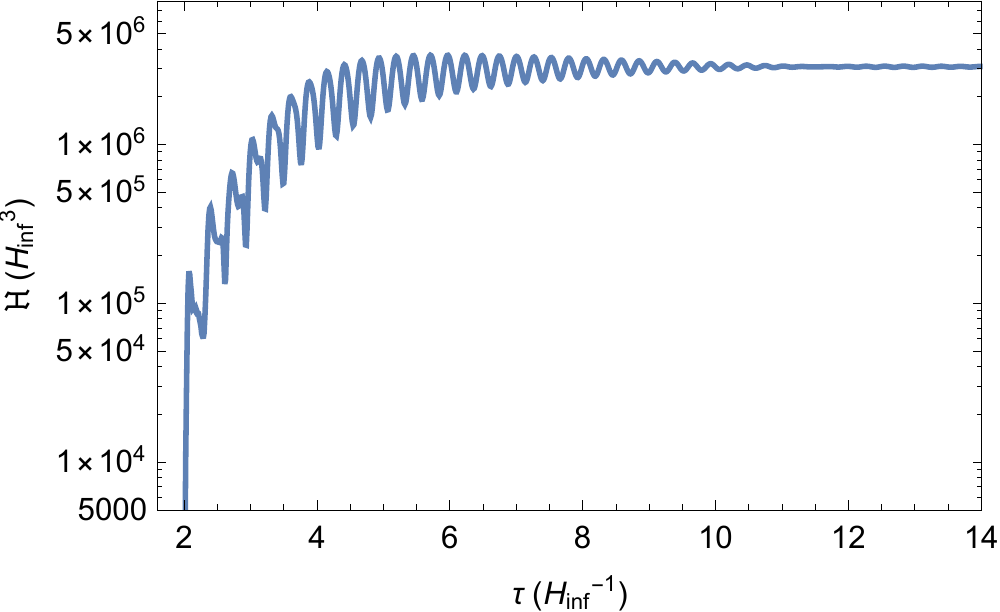}
\end{subfigure}%
\begin{subfigure}{.5\textwidth}
  \centering
  \includegraphics[width=7cm]{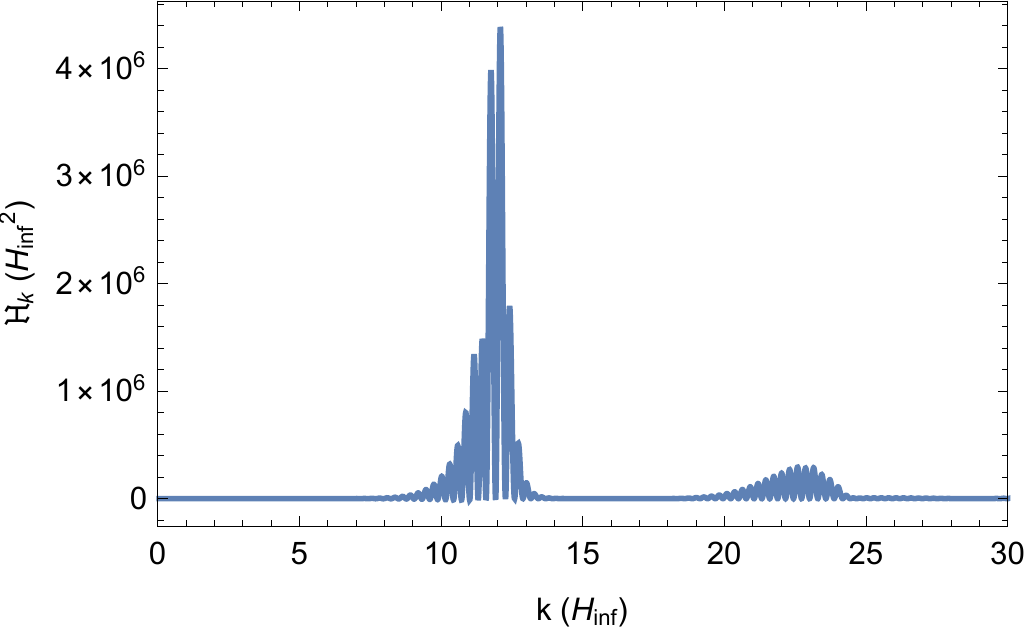}
  \end{subfigure}\\[1ex]
  \begin{subfigure}{.5\textwidth}
  \centering
  \includegraphics[width=7cm]{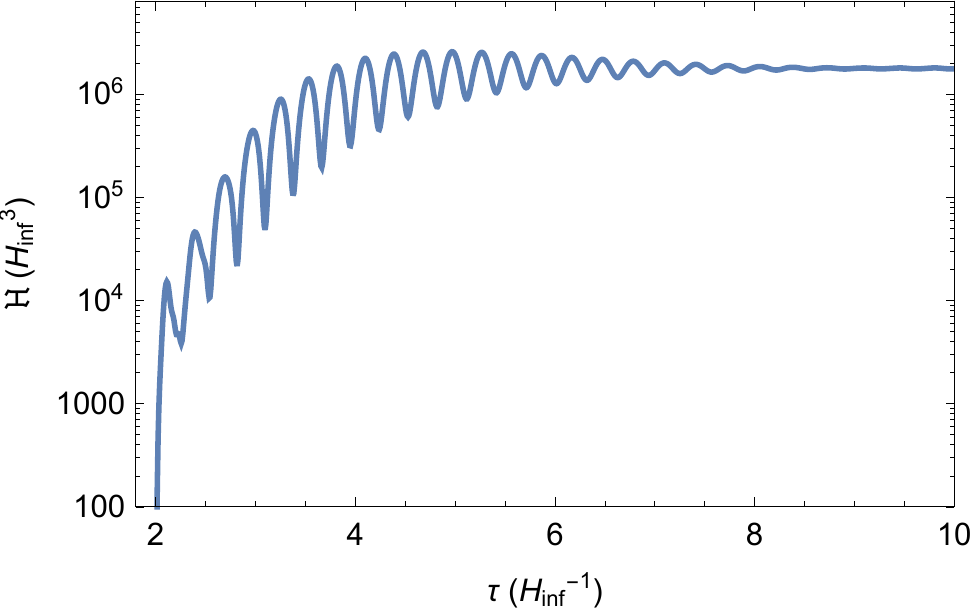}
\end{subfigure}%
\begin{subfigure}{.5\textwidth}
  \centering
  \includegraphics[width=7cm]{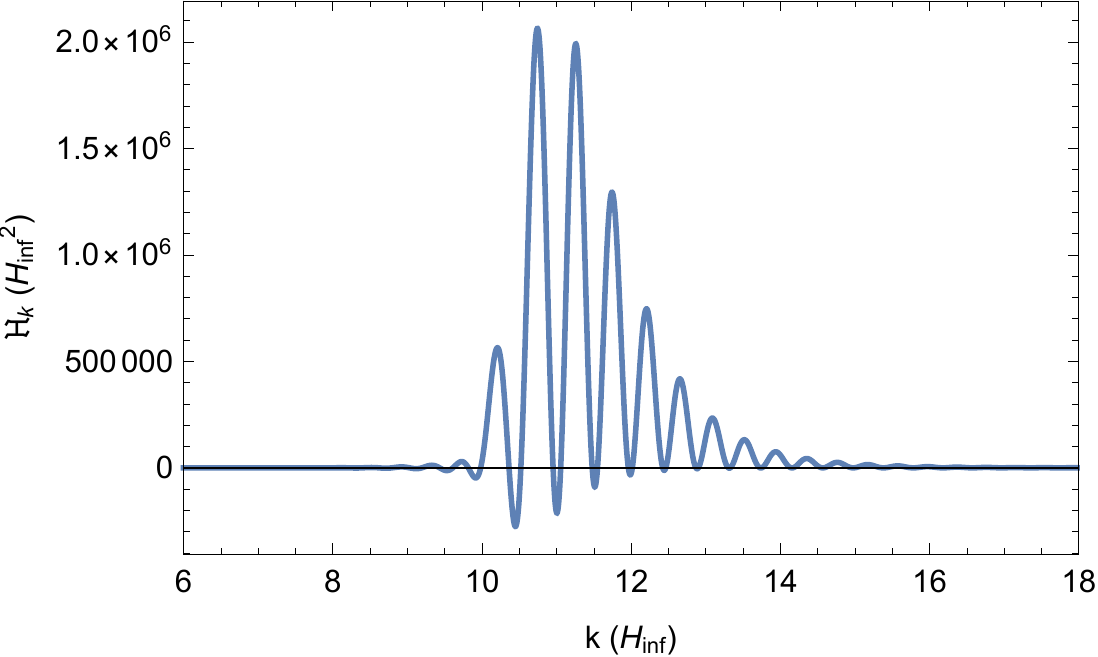}
  \end{subfigure}
\caption{\it The left panels show the time evolution of the helicity $\mathcal{H}$ and the right panels the helicity spectrum $\mathcal{H}_{k}$, i.e.~the integrand in eq.~\eqref{eq:Helicity}, evaluated at the maximal $\tau$ shown in the left panel. 
The first, second and third row is for point 1, 2 and 3 in table~\ref{table:BenchmarkPointsDefinition}, respectively.}
\label{fig:helicity_hk}
\end{figure}

We find that $\lambda_h + \beta_{\lambda_h}/(64 \pi^2)$ in eq.~\eqref{HiggsPotentialMinimum} cannot become sufficiently small to obtain large enough $h_{\rm min} \gg H_{\rm inf}$ with $\xi_R \lesssim 1$. Therefore we need $\xi_R \gg 1$.  As has been pointed out in \cite{Barbon:2009ya}, the coupling between the Higgs and the Ricci scalar lowers the cutoff of the theory to $\Lambda_c \sim M_{\rm Pl} / \xi_R$ for $\xi_R \gg 1$. We will therefore demand that $h_{\rm min} \lesssim M_{\rm Pl}/\xi_R$ during inflation.\footnote{The parameter $\xi_R$ can also be constrained from measurements of the Higgs couplings but the resulting limit is quite weak, $|\xi_R| < 2.6 \cdot 10^{15}$ \cite{Atkins:2012yn}.}  The Higgs VEV contributes to the Planck mass via its coupling to the Ricci scalar. The resulting effective Planck mass is given by $M_{\rm Pl,eff}^2 = M_{\rm Pl}^2 + \xi_R h^2$. For $h \lesssim M_{\rm Pl}/\xi_R$ and $\xi_R\gg 1$, we then have $M_{\rm Pl,eff} \simeq M_{\rm Pl}$.

We have calculated the evolution of the Higgs and the produced gauge fields for the three benchmark points defined in table~\ref{table:BenchmarkPointsDefinition}. Notice that the initial value of the Higgs is a factor $\sim 2 -3$ larger than the scale $M$. One may be worried about the validity of the effective field theory (EFT). As we show in appendix~\ref{appendix:UVcompletion}, however, UV completions exits where higher-dimensional operators of the form $h^{2n} F_{\mu \nu} \tilde{F}^{\mu \nu}$ for $n>1$ are suppressed compared to the leading term with $n=1$ by powers of a small parameter ($\lambda_{Sh}$ in our example). Choosing this parameter small enough ensures the validity of the EFT. We plot the time evolution of the Higgs $h$ and the instability parameter $\xi$ in the EOM of the gauge field \eqref{PhotonEOMk} in fig.~\ref{fig:xih}.
For point 1 (upper-right panel), it is clear from the plot that a net helicity is produced since the instability parameter almost always stays positive, leading to the dominant production of modes with positive helicity. For point 2 and 3 (lower-left and lower-right panel), on the other hand, the instability parameter frequently changes sign but decreases over time. This results in an excess of modes with positive helicity over those with negative helicity. We show the time evolution of the helicity $\mathcal{H}$ in fig.~\ref{fig:helicity_hk}, together with the helicity spectrum $\mathcal{H}_{k}$, i.e.~the integrand in eq.~\eqref{eq:Helicity}, evaluated at the endpoint of the shown time evolution. 
That the (comoving) helicity $\mathcal{H}$ becomes constant follows from the fact that the produced gauge fields evolve as free fields once the production switches off for $\xi \approx 0$. Then $\bm{E} \cdot \bm{B} = 0$ and $\partial_\tau \mathcal{H}=0$ as follows from eq.~\eqref{eq:HelicityEvolution}.

The results for several quantities of interest are summarized in table~\ref{table:BenchmarkPointsResults}. The baryon asymmetry is calculated assuming that the helicity survives until the EW phase transition. Its range arises from the uncertainties in the dynamics of the EW phase transition. We have fixed the reheating temperature for the different points such that the observed value $\eta_{B,{\rm obs}}\simeq 9 \cdot 10^{-11}$ is the upper limit for point 1 and the lower limit for point 2 and 3. The baryon asymmetry increases with the reheating temperature. Therefore the chosen reheating temperature is the lowest one which is consistent with the observed baryon asymmetry (given the uncertainties from the EW phase transition) for point 1 and the highest one for point 2 and 3. The criterion on the CPI temperature is satisfied for all points too. We find that the criterion on the magnetic Reynolds number is more difficult to fulfill though. For point 2 and 3, we find $R_m \sim 1$ at reheating and therefore expect that an inverse cascade develops in the evolution of the magnetic field. Together with the absence of the CPI, this should allow the helicity to survive until the EW phase transition. Note that if MHD simulations find that a somewhat larger magnetic Reynolds number is needed to ensure the inverse cascade, this can be obtained by increasing the parameter $\xi_R$. Decreasing the reheating temperature, on the other hand, would lower the magnetic Reynolds number. For point 1, the magnetic Reynolds number as calculated using the estimate for the velocity eq.~\eqref{eq:EstimateVelocity} is very small, $R_m \sim 3 \cdot 10^{-3}$. This likely means that the inverse cascade does not develop and that the helicity does not survive. Note, however, that eq.~\eqref{eq:EstimateVelocity} has to our knowledge not been numerically verified for our case of diffusion damping. We therefore also give the upper bound $R_m^{\rm max}$ on the magnetic Reynolds number following from energy conservation. We find that $R_m^{\rm max} \sim 1$ for point 1. If the velocity turns out to be sufficiently larger than the estimate eq.~\eqref{eq:EstimateVelocity}, the helicity could survive in this case too. This would also allow us to decrease the parameter $\xi_R$ for point 2 and 3.  A dedicated MHD simulation is clearly desirable to settle this question.

\begin{table}[t]
\centering 
\renewcommand{\arraystretch}{1.3}
\begin{tabular}{| c | c | c | c | c | c | c | c |} 
\hline
 point &  $\mathcal{H}/H_{\rm inf}^3$ & $\rho_{B}/H_{\rm inf}^4$ & $\lambda_{B} H_{\rm inf}$ &  $R_m $  & $R_m^{\rm max}$ & $\hat{T}_{\rm CPI} / \text{GeV}$ & $\eta_{B} $ \\ 
\hline 
 1  &  $5.3 \cdot 10^3$ & $5.6 \cdot 10^4$ & $0.34$  & $3 \cdot 10^{-3}$ & $2$ & $4 \cdot 10^{-3}$ & $(2 \cdot 10^{-13}, 9 \cdot 10^{-11})$\\ 
 2  &  $3.1 \cdot 10^6$ & $2.4 \cdot 10^7$ & $0.41$  & $2$ & $40$ & $1000$ & $(9 \cdot 10^{-11}, 5 \cdot 10^{-8})$\\
 3 &  $1.8 \cdot 10^6$ & $1.2 \cdot 10^7$ & $0.54$  & $2$ & $50$ & $1000$ & $(9 \cdot 10^{-11}, 5 \cdot 10^{-8})$\\
\hline
\end{tabular}
\caption{
{\it Results for the three benchmark points defined in table~\ref{table:BenchmarkPointsDefinition}.
The range for the baryon asymmetry arises due to the uncertainties in the dynamics of the EW phase transition. Due to the uncertainties in our estimates for the magnetic Reynolds number, the CPI temperature and the baryon-to-entropy ratio, we have rounded these quantities to the first significant digit.} } 
\label{table:BenchmarkPointsResults} 
\end{table}

Notice that we have so far worked in the Jordan frame. Going to the Einstein frame removes the coupling of the Higgs to the Ricci scalar and in particular transforms the inflaton potential as
\be
V_{\rm inf} \quad \rightarrow \quad \frac{V_{\rm inf}}{(1 + \xi_R h^2/M_{\rm Pl}^2 )^2} \, .
\ee 
Expanding this for $h\ll M_{\rm Pl}/\sqrt{\xi_R}$ and using $V_{\rm inf}= 3 M_{\rm Pl}^2 H^2$ gives a contribution to the Higgs mass parameter of $\Delta m_h^2=-12 \xi_R H^2$ during inflation, consistent with what we have obtained in the Jordan frame. This also yields a coupling between the Higgs and the inflaton, $\mathcal L  =  2  \xi_R h^2 V_{\rm inf}/ M_{\rm Pl}^2$, which is enhanced by $\xi_R$.
This coupling can mediate decays of the inflaton into SM particles. In particular, the electron mass is $m_e = y_e h / \sqrt{2} \ll H_{\rm inf}$ while the helicity is produced. Assuming an inflaton mass of order $H_{\rm inf}$, the inflaton can decay into electrons already during this stage. As discussed in sec.~\ref{sec:backreaction}, the resulting charged plasma may then backreact on the production of the helicity. 
We show in appendix~\ref{appendix:InflatonDecays}, however, that this backreaction can be expected to be negligible. Furthermore, the transformation to the Einstein frame gives rise to a noncanonical kinetic term for the Higgs doublet.  
This can lead to efficient production of $W^\pm$ and $Z$ bosons while the Higgs is oscillating in its potential as discussed in 
\cite{DeCross:2015uza,Ema:2016dny,Sfakianakis:2018lzf}. Since we always have $ h \ll M_{\rm Pl} / \xi_R$, however, we expect this production not to be enhanced compared to the process already discussed in sec.~\ref{sec:backreaction}.

The magnetic field that arises after the EW phase transition may survive until late times and contribute to the magnetic field which permeates the voids of the universe. The observed $\gamma$-ray spectra of blazars are suppressed in the range of $1-100\;$GeV compared to expectations (see \cite{Durrer:2013pga} for a review). This part of the $\gamma$-ray spectra receives secondary contributions from high-energy $\gamma$-rays which scatter on low-energy photons in the intergalactic voids and produce electron-positron pairs. The latter subsequently upscatter CMB photons via inverse Compton scattering. The flux near the earth from this secondary emission can be suppressed if the intermediate electron-positron pairs are deflected by a magnetic field in the intergalactic voids \cite{Neronov_2010,Tavecchio_2011}. It is then an interesting question if the late-time remnant of the magnetic field that was produced from the relaxing Higgs could explain the observed suppression in the $\gamma$-ray spectra of blazars. In order to answer this, we need to evolve the strength~$B_Y$ of the hypermagnetic field (or $B$ for the magnetic field after the EW phase transition) and its correlation length $\lambda_{B_Y}$ (respectively $\lambda_{B}$) in time until today. The hypermagnetic field initially scales adiabatically and the comoving quantities $B_Y$ and $\lambda_{B_Y}$ are constant. Once enough time has passed for the thermal plasma to affect the hypermagnetic field over its correlation length, $\tau \sim \lambda_{B_Y} / v$ with $v$ being the characteristic plasma velocity, the inverse cascade sets in. We expect that subsequently the hypermagnetic field scales either as in eq.~\eqref{eq:ScalingTurbulence} or eq.~\eqref{eq:ScalingViscous}, depending on the value of the kinetic Reynolds number. At the EW phase transition, the hypermagnetic field is converted into an ordinary magnetic field. We assume that the inverse cascade continues until recombination and that the magnetic field scales adiabatically afterwards until today, i.e.~$B$ and $\lambda_B$ again being constant. Since we find that the kinetic Reynolds number at reheating is much smaller than unity for all benchmark points in table~\ref{table:BenchmarkPointsDefinition}, we expect that the appropriate scaling regime after reheating is given by eq.~\eqref{eq:ScalingViscous}. As one option, we have used this scaling regime for the whole period between reheating and recombination. Since the kinetic Reynolds number grows in time, we have considered as a second option the possibility that the scaling regime changes to eq.~\eqref{eq:ScalingTurbulence} once the kinetic Reynolds number becomes bigger than unity. For completeness, we have as a third option assumed that the scaling regime eq.~\eqref{eq:ScalingTurbulence} applies during the whole period between reheating and recombination. This gives three values for the strength and correlation length of the magnetic field today for each benchmark point in table~\ref{table:BenchmarkPointsDefinition}. We find that for all three options and for each benchmark point the magnetic field strength for the given correlation length is a few orders of magnitude too small to explain the blazar observations \cite{Neronov_2010,Tavecchio_2011}.  As before, we emphasize, however, that this is only a rough estimate for the magnetic field today and that a dedicated MHD simulation would be necessary to establish its size with more confidence. Let us also note that the magnetic field strength that we find is at least 8 orders of magnitude below the upper bound from CMB measurements (see \cite{Durrer:2013pga}). 

The finite correlation length of the hypermagnetic field at the EW phase transition results in baryon isocurvature perturbations.  While the correlation length is typically too small to be constrained by CMB measurements, baryon isocurvature perturbations on much shorter length scales can potentially spoil the success of BBN \cite{Kamada:2020bmb}. In order to check that this does not happen, we have evaluated the correlation length of the hypermagnetic field at the EW phase transition in the three different scaling regimes discussed above.  This subsequently becomes the correlation length of the baryon isocurvature perturbations which we have then adiabatically scaled to the time of BBN. We have found that the resulting correlation length is in all cases many orders of magnitude smaller than the neutron diffusion length, leading to efficient damping of the baryon isocurvature perturbations, and the corresponding constraint \cite{Kamada:2020bmb} is easily satisfied.

\subsection{Higgs potential with coupling to the inflaton}

Finally, let us comment on another option to induce the tachyonic mass term for the Higgs during inflation. Instead of the coupling of the Higgs to the Ricci scalar in eq.~\eqref{eq:RicciScalarCoupling}, we could consider a coupling to the inflaton. The combined potential then reads
\begin{equation}
V(\Phi,\chi)\, =\, -\lambda_{\chi h}\chi^2|\Phi|^2\, + \, \lambda_h|\Phi|^4 \, + \,V_{\rm inf}(\chi) \,,
\label{eq:Higgs-inflaton}
\end{equation}
where $\chi$ is the inflaton, $V_{\rm inf}(\chi)$ the inflationary potential and we neglect the Higgs mass parameter $m_h^2$. This leads to a minimum of the Higgs potential which in terms of the inflaton is given by
\be
h_{\rm min}(\chi)=\sqrt{\frac{\lambda_{\chi h}}{\lambda_h+\beta_{\lambda_h}/(64\pi^2)}}\,\chi \, .
\label{eq:minhiggs}
\ee
Assuming that the Higgs sits in this minimum, the potential for the inflaton is
\be
V(\chi)\, \simeq \, -\frac{1}{4}\lambda_h h_{\rm min}^4(\chi) \, + \, V_{\rm inf}(\chi)\, ,
\ee
where we are neglecting $\beta_{\lambda_h}/(64 \pi^2)$ as compared to $\lambda_h$.
From this, we find the condition 
\be
V_{\rm inf}(\chi) \, > \, \frac{1}{4}\, \lambda_h h_{\rm min}^4(\chi)
\label{eq:condition}
\ee
to ensure that the inflationary potential is not perturbed. 

In particular, condition (\ref{eq:condition}) should be fulfilled at the end of inflation. Denoting the value of the inflaton at this time by $\chi_0$, we have $V_{\rm inf}(\chi_0)=3 M_{\rm Pl}^2 H_{\rm inf}^2$, where $H_{\rm inf}$ is the Hubble parameter at the end of inflation. Taking, as we are doing in this paper, $H_{\rm inf} = 10^{13}\,$GeV and $\lambda_h\simeq 0.005-0.01$ for the relevant scales (cf.~the upper left panel of fig.~\ref{fig:xih}), we obtain the upper bound 
\be
h_{\rm min}(\chi_0) \, \lesssim \, 3 \times  10^{16}\; \textrm{GeV}\
\label{eq:boundhmin}
\ee
on the Higgs value at the end of inflation and thus the initial value $h_{\rm ini}=h_{\rm min}(\chi_0)$ for the later time evolution. Moreover, from eqs.~(\ref{eq:minhiggs}) and (\ref{eq:slow}), and assuming $\chi_0 \approx M_{\rm Pl}$, we get $\dot{h}_{\rm ini}\simeq -h_{\rm min}H_{\rm inf}$ as the initial condition for the time derivative.

Of course condition (\ref{eq:condition}) should be fulfilled for the whole inflationary period. In addition we should also ensure that the joint potential for the Higgs and inflaton has no runaway directions. Both requirements translate into an upper bound on the coupling $\lambda_{\chi h}$. However this bound is model-dependent. Just to illustrate this point we will consider here a simple model of chaotic inflation based on the potential
$V_{\rm inf}=V_2(\chi)+V_4(\chi)$ where $V_2(\chi)=m^2_\chi\chi^2$ and $V_4(\chi)=\lambda_\chi \chi^4$. We will assume that the inflaton potential is dominated during inflation by the quartic potential $V_{\rm inf}(\chi)\simeq V_4(\chi)$, i.e.~for $\chi\in\left[2\sqrt{2},\sqrt{8(1+N)}\right]M_{\rm Pl}$, in which case $\lambda_\chi\simeq 3.4 \times 10^{-14}$ as follows from the Planck measurement of the primordial amplitude
$A_s=2.1\times 10^{-9}$~\cite{Akrami:2018odb} at $N=60$ e-folds before the end of inflation. From eq.~(\ref{eq:condition}), the upper bound on $\lambda_{\chi h}$ then reads
\be
\lambda_{\chi h} \, \lesssim \, 2\sqrt{\lambda_\chi \lambda_h} \, \approx \, 3\times 10^{-8} .
\label{eq:condition2}
\ee 
This condition also guarantees the absence of runaway directions. At the end of inflation, defined as the time when the slow-roll parameter $\epsilon=1$, the inflaton value in this model is $\chi_0=2\sqrt{2} M_{\rm Pl}$. This gives $H_{\rm inf} \simeq 2.1 \times 10^{12}$ for the Hubble rate at the end of inflation, i.e.~somewhat lower than what we have considered in this paper. From this, we get a slightly more stringent condition
\be
h_{\rm min}(\chi_0) \, \lesssim \, 10^{16} \; \textrm{GeV}
\ee
on the Higgs value at the end of inflation.

After inflation ends, the inflaton oscillates in its potential. We assume that at this stage the inflaton potential is dominated by the quadratic term, $V_{\rm inf}(\chi)\simeq V_2(\chi)$, in which case one finds that $m_\chi\simeq 1.3 \times 10^{12}$ GeV. The amplitude of the inflaton then redshifts like $a^{-3/2}$ and the time-average of the 
Higgs mass term $-\lambda_{\chi h}\chi^2$ decreases as $a^{-3}$. This is similar to what happens in the case of the coupling to the Ricci scalar and induces a movement of the Higgs. Note, however, that using the time-averaged value for the Higgs mass term may not be a good approximation here since it changes on a similar timescale $\sim H_{\rm inf}^{-1}$ as the Higgs. It may therefore be necessary to solve for the joint time-evolution of the Higgs and the inflaton. Note also that in this model the Higgs mass term and thus the minimum of the Higgs potential already change significantly during inflation. The resulting movement of the Higgs during inflation may lead to an important additional contribution to photon production. We leave a further study of this model for future work.

\section{Conclusions}
\label{sec:conclusions}

In this paper, we have considered a scenario where the Higgs couples to the Chern-Simons term of the hypercharge gauge group, $\sim |\Phi|^2 Y_{\mu \nu} \tilde{Y}^{\mu \nu}$. If the Higgs is away from the late-time minimum of its potential during inflation, it relaxes toward this minimum after inflation ends. The time-dependent Higgs value then yields a source term for the photon and can result in the production of a net magnetic helicity. At reheating this helicity is converted into hypermagnetic helicity which under certain conditions can survive until the EW phase transition. At this point, the hypermagnetic helicity is converted back into ordinary magnetic helicity. The former enters into the anomaly equation of the current for baryon plus lepton number, $B+L$, while the latter does not. The conversion of the helicity during the EW phase transition therefore leads to the production of a compensating $B+L$-asymmetry. This is partly washed out by EW sphalerons but since the latter freeze-out while the phase transition progresses, a sizeable net $B+L$-asymmetry can remain.

We have first presented the EOMs for the Higgs and the photon which need to be solved in order to determine the produced magnetic helicity. For simplicity, we have neglected the backreaction of photon production onto the Higgs and finite-temperature effects and have discussed the conditions under which this is justified (conditions which we have checked in all presented cases). 
After reheating, the resulting hypermagnetic fields start to interact with the thermal plasma and both evolve jointly afterwards. The comoving helicity can be approximately conserved during this evolution if the magnetic Reynolds number is sufficiently large. Furthermore, fermionic asymmetries are produced together with the helicity and can partly cancel the helicity by a chiral plasma instability if they are not erased by EW sphalerons before. We have estimated the resulting conditions on the helicity, energy density and correlation length of the produced magnetic fields and on the reheating temperature. Subsequently, we have presented an estimate of the baryon asymmetry which is generated from the helicity conversion during the EW crossover. The theoretical uncertainties on these estimates are, however, quite large. Firstly, this is due to the fact that to our knowledge no numerical MHD simulation exists in the regime which is relevant for us, with small kinetic and large magnetic Reynolds number and in the presence of fermionic asymmetries. We have instead used analytical estimates which are less reliable than a full numerical simulation. Additional theoretical uncertainties arise from the dynamics of the EW phase transition. Both types of uncertainties will hopefully shrink in the future though, with dedicated MHD simulations and more lattice studies of the EW crossover. 

We have assumed a simple model to obtain a large, initial Higgs value. To this end, we have included a coupling of the Higgs to the Ricci scalar. The sign was chosen such that it yields a tachyonic mass term for the Higgs during inflation. We have also added a singlet scalar which couples to the Higgs and prevents the Higgs quartic coupling from running to negative values. During inflation, the Higgs potential then has a minimum at large field values. After inflation ends, this minimum and with it the Higgs move toward the origin. Via the coupling to the Chern-Simons term of the hypercharge gauge group, this movement leads to the production of a net helicity. 
We have studied three benchmark points to show that the helicity can survive until the EW phase transition and can reproduce the observed baryon asymmetry of the universe. 
Other models are conceivable to induce a large, initial Higgs value and it would be interesting to explore them. In particular, we have commented on another possibility to obtain a large tachyonic Higgs mass during inflation, by coupling the Higgs to the inflaton. We leave a more detailed study of this case for future work. 

Finally, we wish to stress that Higgs relaxation has already been used in the literature \cite{Kusenko:2014lra,Pearce:2015nga,Yang:2015ida} to generate the baryon asymmetry, although, as we pointed out in sec.~\ref{sec:introduction}, with important differences in the corresponding mechanisms. In both approaches the Higgs is assumed to keep a large VEV during inflation such that its post-inflationary relaxation creates an effective chemical potential. In refs.~\cite{Kusenko:2014lra,Pearce:2015nga,Yang:2015ida} this is done by assuming higher-dimensional operators with appropriately chosen Wilson coefficients, and in our approach in this paper we have produced such an effect by a coupling of the Higgs field to the Ricci scalar or to the inflaton, with similar results in all cases. 
From here on both mechanisms are different. In refs.~\cite{Kusenko:2014lra,Pearce:2015nga,Yang:2015ida} lepton number is explicitly broken by the Majorana masses of right-handed neutrinos and a lepton asymmetry is generated at finite temperature since the time-variation of the Higgs field in the relaxation phase creates an effective chemical potential for lepton number. This is then redistributed by weak sphalerons into a lepton and baryon number asymmetry. However in our mechanism, the time variation of the relaxing Higgs field is a source term for the photon and can result in the production of net magnetic helicity, which is in turn transformed into a $B+L$-asymmetry at the electroweak phase transition. In order to ensure efficient production of magnetic helicity, we have assumed that this process takes place before reheating. Since the asymmetry is ``stored'' in the hypermagnetic helicity until the electroweak phase transition, our mechanism does not need any explicit source of $B-L$-violation, so in particular it does not rely on the structure of Majorana right-handed neutrino masses. In a way one could say that both mechanisms are complementary, as one could imagine a combination of both: i.e.~a mechanism where helical magnetic fields could be generated during Higgs relaxation before reheating, and a lepton asymmetry is produced in the presence of lepton-number violation if the Higgs continues to relax after reheating.

 \section*{Acknowledgements}
We would like to thank Valerie Domcke and Evangelos Sfakianakis for very useful discussions and comments. 
This work is supported by the Secretaria d'Universitats i Recerca del Departament d'Empresa i Coneixement de la Generalitat de Catalunya under the grant 2017SGR1069, by the Ministerio de Economía, Industria y Competitividad under the grant FPA2017-88915-P and from the Centro de Excelencia Severo Ochoa under the grant SEV-2016-0588. 
IFAE is partially funded by the CERCA program of the Generalitat de Catalunya. YC is supported by the European Union's Horizon 2020 research and innovation programme under the Marie Sklodowska-Curie grant agreement No.~754558. 
 
\appendix

\section{Ultraviolet completion}
\label{appendix:UVcompletion}

In this appendix, we present a possible UV completion which gives rise to the dimension-six operator
\be
\mathcal L \, = \, \frac{1}{2}\frac{|\Phi|^2}{M^2}Y_{\mu \nu}\tilde Y^{\mu\nu}
\label{eq:lagrangiano}
\ee
that we are using to generate the magnetic field after the end of the inflationary period. 

A very simple model consists of a complex scalar field $S$, a singlet under the SM gauge group, which interacts with  $Y_{\mu \nu}\tilde Y^{\mu\nu}$ via the dimension-five operator
\be
\mathcal L \, = \, \frac{1}{2f_S}\left( e^{i\alpha}S \, + \, \text{h.c.} \right)\,Y_{\mu \nu}\tilde Y^{\mu\nu} ,
\label{eq:SFFtilde}
\ee
where $\alpha$ is an arbitrary phase and $f_S$ is a mass scale. After decomposition of the complex field into its real and imaginary parts, $S=r+i a$, this gives rise to the usual axial coupling $(a/f_S)Y_{\mu \nu}\tilde Y^{\mu\nu}$ for $\alpha=\pi/2$, but of course the coupling can be much more general.

We will now consider a general renormalizable potential for the field $S$, with a coupling to the Higgs doublet $\Phi$ as
\be
V(S,\Phi) \, = \, -\mu(e^{i\alpha}S  +\text{h.c.})|\Phi|^2+m_S^2 |S|^2 \, + \, \lambda_{Sh}|S|^2| \Phi|^2 \, + \, \frac{1}{2}\lambda_S^2|S|^4 \, + \, V_{\rm SM}(\Phi) \, ,
\label{eq:potential}
\ee
where $\mu\geq 0$ is a mass parameter, $m_S^2 \geq 0$ is the (common) mass-squared of the real and imaginary parts of $S$, and $\lambda_{Sh}\geq 0,\,\lambda_S$ are real couplings. The global invariance $S\to e^{i \theta_S}S$, where $\theta_S$ is an arbitrary phase, is explicitly broken by the first term in the potential (\ref{eq:potential}) which then prevents the appearance of a massless Goldstone boson if $S$ acquires a VEV once EW symmetry is broken. 

For momenta much smaller than $m_S$, the field $S$ is decoupled from the theory and can be integrated out neglecting its kinetic term and simply using its potential. Minimization of the potential (\ref{eq:potential}) yields for 
\be
S=e^{i \theta}|S|
\ee
the equations determining the minimum
\begin{align}
\theta& \, = \, -\alpha \nonumber\\
\mu |\Phi|^2& \,= \, (m_S^2+\lambda_{Sh}|\Phi|^2+\lambda_S^2|S|^2) \, |S| \, .
\label{eq:eom}
\end{align}

A quick glance at eq.~(\ref{eq:eom}) shows that for $\mu=0$ the only solution is $|S|=0$. For $\mu\neq 0$ and $\lambda_S,m_S\neq 0$ the solution, as can be seen from eq.~(\ref{eq:eom}), depends on two parameters,
\be
\frac{\lambda_S |S|}{m_S}=f(x^2,y^2),\quad \;\, y^2\equiv \lambda_S\frac{\mu}{m_S}\frac{|\Phi|^2}{m_S^2}, \quad \; \, x^2
\equiv\lambda_{Sh}\frac{|\Phi|^2}{m_S^2},
\ee
which can be solved analytically. For the validity of the EFT expansion, the parameters $x$ and $y$ should be small. 
A power series expansion in $x$ and $y$ then gives
\begin{align}
|S|&\, = \,\mu \frac{|\Phi|^2}{m_S^2}\left[\frac{1}{1+x^2}-\frac{y^4}{(1+x^2)^4}+\frac{3y^8}{(1+x^2)^7}+\cdots \right]\nonumber\\
& \, = \,\mu \frac{|\Phi|^2}{m_S^2}\left[1-\lambda_{Sh} \frac{|\Phi|^2}{m_S^2}+\mathcal O(|\Phi|^4/m_S^4) \right],
\label{eq:expansionS}
\end{align}
and the term in eq.~\eqref{eq:SFFtilde} at the minimum yields
\begin{align}
\mathcal L & \, = \, \frac{\mu}{f_S} \frac{|\Phi|^2}{m_S^2}\left[\frac{1}{1+x^2}-\frac{y^4}{(1+x^2)^4}+\frac{3y^8}{(1+x^2)^7}+\cdots \right]\,Y_{\mu \nu}\tilde Y^{\mu\nu} \nonumber\\
& \, = \,  \frac{\mu}{f_S} \frac{|\Phi|^2}{m_S^2}\left[1-\lambda_{Sh} \frac{|\Phi|^2}{m_S^2}+\mathcal O(|\Phi|^4/m_S^4) \right]\,Y_{\mu \nu}\tilde Y^{\mu\nu} \, .
\label{eq:SFFtildeS}
\end{align}
Matching the leading term with eq.~(\ref{eq:lagrangiano}) we find
\be
\label{Mrelation}
M=\sqrt{\frac{f_S}{2 \mu}} \, m_S \, .
\ee
Similarly, the potential (\ref{eq:potential}) at the minimum is given by
\begin{align}
V(\Phi)&=\frac{\mu^2}{m_S^2}|\Phi|^4\left[- \frac{1}{1+x^2}+\frac{y^4}{2(1+x^2)^4}-\frac{y^8}{(1+x^2)^7} +\cdots  \right] \, + \, V_{\rm SM}(\Phi)\nonumber\\
&=\frac{\mu^2}{m_S^2}|\Phi|^4\left[-1+\lambda_{Sh } \frac{|\Phi|^2}{m_S^2}+\mathcal O(|\Phi|^4/m_S^4)\right] \, + \, V_{\rm SM}(\Phi) \,.
\label{eq:potential-exp}
\end{align}

Consistent with the condition $x^2,y^2\ll1$, we will consider
%
field configurations such that 
\be
\frac{\lambda_{Sh} f_S}{2 \mu}|\Phi|^2\ll M^2,\quad \; \frac{\lambda_S^2 f_S^3}{8 \mu} |\Phi|^4\ll M^6 \, ,
\ee
where we have used eq.~\eqref{Mrelation}. We are interested in field values up to $|\Phi| \sim M$. The conditions can then be fulfilled for example for $\lambda_{Sh} \ll 1$ and $\mu\approx f_S\ll M$.
This ensures that higher-dimensional operators in eq.~\eqref{eq:SFFtildeS} of the form $|\Phi|^{2n} Y_{\mu\nu}\tilde Y^{\mu\nu}$ for $n\geq 2$ and corrections to $V_{\rm SM}(\Phi)$ in eq.~\eqref{eq:potential-exp} are greatly suppressed.

Finally, note that a simple way of generating a term like that in eq.~(\ref{eq:SFFtilde}) is through the introduction of a massive hypercharged vector-like  (Dirac) fermion $\chi$ with Yukawa coupling to $S$~\cite{Carena:2019xrr} $\lambda=|\lambda|e^{i\theta_\lambda}$, where $\theta_\lambda$ is an arbitrary phase. The corresponding term reads
\be
\mathcal L
\, =  \, \lambda \, \bar\chi_L S \chi_R + \text{h.c.} \, = \, |\lambda| |S|\left[\cos(\theta_\lambda-\alpha)\bar\chi\chi + \sin(\theta_\lambda-\alpha)\bar\chi i\gamma_5\chi\right] ,
\label{eq:interaccion}
\ee
where the EOMs (\ref{eq:eom}) for the field $S$ have been used in the second step.
For the phase values $\theta_\lambda=\alpha\pm\pi/2$, eq.~(\ref{eq:interaccion}) yields
\be
\mathcal L \, = \, \pm|\lambda| |S|\bar\chi i\gamma_5\chi \, .
\ee
Through one-loop diagrams where a loop of $\chi$-fermions is exchanged and emits two photons, this gives rise to the interaction in eq.~(\ref{eq:SFFtilde}) evaluated in the minimum in eq.~\eqref{eq:eom}.\footnote{For arbitrary values of the phase $\theta_\lambda$, the coefficient of the term $\bar\chi\chi$ in eq.~(\ref{eq:interaccion}) does not vanish, and the corresponding interaction would also give rise to the Lagrangian term $|S| Y_{\mu\nu}Y^{\mu\nu}$.}

\section{Solving the electroweak vacuum instability}
\label{appendix:RGEs}

In this appendix, we will present a possible SM completion to avoid the instability of the SM vacuum. The simplest possibility is to introduce a massive real singlet $\phi$ which couples to the Higgs by a quartic coupling. The joint potential reads
\be
V(\phi,\Phi)=\frac{1}{2} m_\phi^2\phi^2+\lambda_{\phi h}|\Phi|^2\phi^2+\lambda_\phi \phi^4+V_{\rm SM}(\Phi) \, ,
\ee
where we have introduced the symmetry $\phi\to-\phi$. The coupling $\lambda_{\phi h}$ will contribute positively to the $\beta$-function of the Higgs quartic coupling for scales $\mu>m_\phi$ and can avoid the SM instability, a phenomenon which mainly depends on the values of $m_\phi$ and of $\lambda_{\phi h}$.  

The RG equations for all dimensionless parameters ($X$), including those of the SM, are
\be
\frac{d X}{dt}\equiv \frac{1}{16\pi^2}\beta_X\, , 
\ee 
where $t=\log\mu$ and $\mu$ is the renormalization scale. With $t_0=\log m_\phi$, the $\beta$-functions are given by~\cite{Gonderinger:2009jp}
\begin{align}
\label{betafunctions}
\beta_{g_1}&=\frac{41}{10}g_1^3\, ,\quad \; \, \beta_{g_2}=-\frac{19}{6}g_2^3 \, ,\quad \; \, \beta_{g_3}=-7 g_3^3\, ,\nonumber\\
\beta_{y_t}&=y_t\left( \frac{9}{2}y_t^2-8 g_3^2-\frac{9}{4}g_2^2-\frac{17}{20}g_1^2  \right) ,\nonumber\\
\beta_{\lambda_{h}}&=\lambda_h \left(24\lambda_h+12 y_t^2-9 g_2^2-\frac{9}{5}g_1^2  \right)-6 y_t^4+\frac{27}{200}g_1^4+\frac{9}{20}g_1^2 g_2^2+\frac{9}{8} g_2^4 
+2\lambda_{\phi h}^2\theta(t-t_0)\, , \nonumber\\
\beta_{\lambda_{\phi h}}&=\left[12\lambda_h\lambda_{\phi h}+8\lambda_{\phi h}^2+24\lambda_{\phi h}\lambda_\phi-\lambda_{\phi h}\left(
\frac{9}{2}g_2^2+\frac{9}{10} g_1^2-6 y_t^2 \right)\right]\theta(t-t_0)\, ,\nonumber\\
\beta_{\lambda_\phi}&=\left[ 2 \lambda_{\phi h}^2+72 \lambda_\phi^2  \right]\theta(t-t_0) \, .
\end{align}

\section{The Higgs potential during inflation}
\label{appendix:HiggsPotentialInflation}

In this appendix, we provide some details about the Higgs potential during the inflationary period. We assume a coupling of the Higgs $h$ to the Ricci scalar $R$:
\be
\mathcal L=-\frac{1}{2}\xi_R R \, h^2  .
\ee 
The Ricci scalar for a flat universe is given in terms of the scale factor $a$ by
\be
R=-6\left[\frac{\ddot a}{a}+\left(\frac{\dot a}{a}\right)^2  \right] ,
\ee
where $\dot a\equiv da/dt$ with $t$ being the cosmological time. In terms of the slow-roll parameter $\epsilon$ during inflation, ${\ddot a}$ can be expressed as 
\be
\frac{\ddot a}{a}=[1-\epsilon(\chi)]\, H^2 ,
\label{eq:epsilon}
\ee
where $\chi$ is the inflaton and $H=\dot a/a$ the Hubble parameter. Using this, we can write
\be
R=-6 \, [2-\epsilon(\chi) ] \,H^2 .
\ee

Let us denote the value of the inflaton (Hubble parameter) for $N$ e-folds before the end of inflation as $\chi_N$  ($H_N$). At $N=60$, one has $\epsilon(\chi_{60})\ll1$ and thus $R\simeq-12 H_{60}^2$. The nominal end of inflation, i.e.~$N=0$, occurs in models with a single inflaton for $\epsilon(\chi_0)\equiv 1$. This gives $R \simeq-6 H_0^2$. In multi-field models like hybrid inflation, on the other hand, one can have $\epsilon(\chi_0)\ll 1$ and $R \simeq-12 H_0^2$. For the subsequent period of matter domination, $a\propto t^{2/3}$ and $R=-3 H_0^2/a^{3}$. Note that we denote the Hubble rate at the end of inflation by $H_{\rm inf}$ in the rest of the paper, i.e.~$H_0 = H_{\rm inf}$.

The Higgs potential during inflation is given by
\be
V \,= \,-\frac{6 [2-\epsilon(\chi_N)]}{2} \xi_R H_N^2 h_N^2+\frac{1}{4}\lambda_h h_N^4 \, ,
\ee
where $h_N$ is the value of the Higgs at $N$ e-folds before the end of inflation. The Higgs potential has a minimum at
\be
h_N \, = \, \sqrt{\frac{6 [2-\epsilon(\chi_N)] \xi_R}{\lambda_h+\beta_{\lambda_h}/(64 \pi^2)}}H_N  \,.
\label{eq:hN}
\ee
Using the slow-roll equations for the inflaton
\be
\epsilon(\chi)=\frac{1}{2 M_{\rm Pl}^2}\left( \frac{\dot \chi}{H} \right)^2=\frac{M_{\rm Pl}^2}{2}\left(\frac{V_{\rm inf}'(\chi)}{V_{\rm inf}(\chi)}\right)^2 ,
\label{eq:slow}
\ee
and that the inflationary potential is $V_{\rm inf}(\chi)=3 H^2 M_{\rm Pl}^2$, we find
\be
\dot H_N=-\epsilon(\chi_N)\, H_N^2 \, .
\label{Hdot}
\ee 
Taking the time derivative of $h_N$ in eq.~(\ref{eq:hN}), neglecting the field dependence in the square-root and using eq.~(\ref{Hdot}) we then get
\be
\dot h_N=- \epsilon(\chi_N)h_N H_N \, .
\label{eq:velocity}
\ee

In models with a single inflaton, the rolling of the inflaton triggers the breakdown of the slow-roll conditions at some value $\chi=\chi_0$ such that $\epsilon(\chi_0)=1$ and the value and velocity of the Higgs at the end of inflation are
\be
h_0\, = \, \sqrt{\frac{6\, \xi_R}{\lambda_h+\beta_{\lambda_h}/(64 \pi^2)}} \, H_0,\quad \quad \dot h_{0} \, = \,- h_{0}H_0 \, .
\ee
This sets the initial conditions of the Higgs at the onset of matter domination, if the transition to this era from the inflationary period is very fast.

However, in multi-field models or hybrid inflation~\cite{Linde:1993cn}, there is an instability of the extra (waterfall) field which is triggered when the inflaton reaches a critical value $\chi_c$, with corresponding Hubble parameter $H_c$. If such a value is reached when $\epsilon\ll 1$, then the value and velocity of the Higgs at the end of inflation are
\be
h_c=\sqrt{\frac{12\xi_R}{\lambda_h+\beta_{\lambda_h}/(64 \pi^2)}}H_c,\quad \quad \dot h_c=0 \, .
\ee
For a very fast transition from the inflationary to the matter-dominated era, this again sets the initial conditions for the Higgs at the onset of matter domination.

\section{Inflaton decays into standard model particles}
\label{appendix:InflatonDecays}

As we have discussed in sec.~\ref{sec:HiggsPotential}, after going to the Einstein frame the nonminimal coupling eq.~\eqref{eq:RicciScalarCoupling} gives rise to a coupling between the Higgs and the inflaton, 
\be
\label{InflatonHiggsCoupling}
\mathcal L\, =\, 2\xi_R \frac{h^2}{M_{\rm Pl}^2}V_{\rm inf}\, ,
\ee
which is enhanced by $\xi_R$. 
Let us decompose the Higgs and inflaton into  quantum components ($h,\chi$) and classical components ($h_c,\chi_c$) as $h\to h+h_c$, $\chi\to\chi+\chi_c$. Expanding to first order in the quantum components for each field and using eq.~\eqref{eq:slow}, we find a mixing between the Higgs and the inflaton at the end of inflation (i.e.~for $\epsilon=1$) of 
\be
\label{HiggsInflatonMixing}
\delta_{\chi h}=4\sqrt{2}\xi_R \frac{H_{\rm inf}^2}{\lambda_h h_c M_{\rm Pl}}  \, .
\ee
Now considering typical values for the parameters, $\xi_R=10^2$, $H_{\rm inf}=10^{13}\,$GeV, $\lambda_h=0.01$ and $h_c=10^{15}\,$GeV, we obtain
\be
\delta_{\chi h}\, \simeq \, 2\times 10^{-3}
\ee
at the end of inflation. Since the inflaton potential decreases with the expansion of the universe, the mixing subsequently becomes smaller too. 
 
We can now obtain the total decay rate of the inflaton into SM particles $\Gamma_{\chi\to \rm SM}$ by rescaling the corresponding rate for the Higgs in the SM, $\Gamma_{h}=0.004$ GeV, and taking into account its mixing with the inflaton. This gives 
\be
\label{InflatonDecayRate}
\Gamma_{\chi\to \rm SM} \, = \, \delta_{\chi h}^2 \frac{m_\chi}{m_h}\, \Gamma_h \, \simeq \, 2 \times 10^{-8}\, \textrm{GeV} \, \frac{m_\chi}{m_h} \, ,
\ee
where $m_h\simeq 125\,$GeV is the Higgs mass today. The condition in eq.~\eqref{PlasmaInteractionCondition} to ensure that the charged plasma does not backreact on the production of the helicity is then fulfilled for inflaton masses $m_\chi \gtrsim 3 \times 10^{11} \,$GeV. This can certainly be satisfied. Furthermore, note that this bound on the inflaton mass is conservative. Firstly, it was derived assuming a constant Higgs-inflaton mixing equal to its value at the end of inflation in eq.~\eqref{HiggsInflatonMixing}. However, the mixing actually decreases with the expansion of the universe as discussed above. Secondly, even at the end of inflation the inflaton decay rate in eq.~\eqref{InflatonDecayRate} is an overestimate since many decay channels are kinematically blocked due to the large Higgs VEV at this time. For example, if we just consider the decay into some light charged particles, $\chi\to\ell c $ with $\ell c=e^+e^-,\mu^+\mu^-,u\bar u, d\bar d,s\bar s$, and using that 
${\rm BR}(h\to\ell c) \simeq 7.5\times 10^{-4}$ we get
\be
\Gamma_{\chi\to \ell c} \, \simeq \, 2 \times 10^{-11} \, \textrm{GeV} \, \frac{m_\chi}{m_h} \, .
\ee
This would give the bound $m_\chi \gtrsim 9 \times 10^{9}$ GeV on the inflaton mass. For a large Higgs VEV, even less channels may be kinematically accessible, making the bound even weaker.

\bibliographystyle{JHEP}
\bibliography{Biblio_HiggsBaryo}

\end{document}